yn%\documentclass[aps,prd,twocolumn,groupedaddress]{revtex4-1}
\documentclass[aps,prd,twocolumn]{revtex4}

\usepackage{aas_macros}
\usepackage{natbib}

\usepackage[latin1]{inputenc}
\usepackage{amsfonts}
\usepackage{amsmath}
\usepackage[T1]{fontenc}
\usepackage[dvips]{graphicx}

\newcommand{\be}{\begin{equation}}
\newcommand{\ee}{\end{equation}}
\newcommand{\ba}{\begin{eqnarray}}
\newcommand{\ea}{\end{eqnarray}}
\newcommand{\ch}{\mathcal{H}}
\newcommand{\qc}{$Q\| u_c~$}
\newcommand{\qx}{$Q\| u_x~$}

\begin{document}

\title{Interacting Dark Energy -- constraints and degeneracies}

\author{Timothy Clemson$^1$, Kazuya Koyama$^1$, Gong-Bo Zhao$^1$, Roy Maartens$^{1,2}$, Jussi V\"aliviita$^3$}

\affiliation{$^1$Institute of Cosmology \& Gravitation, University of Portsmouth, Portsmouth PO1 3FX, United Kingdom}

\affiliation{$^2$Department of Physics, University of Western Cape, Cape Town 7535, South Africa}

\affiliation{$^3$Institute of Theoretical Astrophysics, University of Oslo, N-0315 Oslo, Norway}

\date{\today}

\begin{abstract}

In standard cosmologies, dark energy interacts only gravitationally with dark matter. There could be a non-gravitational interaction in the dark
sector, leading to changes in the effective DE equation of state, in the redshift dependence of the DM density and in structure formation.
We use CMB, BAO and SNIa data to constrain a model where the energy transfer in the dark sector is proportional to the DE density. There are
two subclasses, defined by the vanishing of momentum transfer either in the DM or the DE frame. We conduct a Markov-Chain
Monte-Carlo analysis to obtain best-fit parameters. The background evolution allows large interaction strengths, and the constraints from CMB
anisotropies are weak. The growth of DM density perturbations is much more sensitive to the interaction, and can deviate strongly from the standard
case. However, the deviations are degenerate with galaxy bias and thus more difficult to constrain. Interestingly, the ISW signature is suppressed
since the non-standard background evolution can compensate for high growth rates. We also discuss the partial degeneracy between interacting DE and
modified gravity, and how this can be broken.

\end{abstract}

\maketitle

\section{Introduction}

The late-time acceleration of the expansion of the Universe demands explanation and observational verification. Currently observational tests of the
standard $\Lambda$CDM cosmological model are not  precise enough to adequately rule out the wide variety of alternative
dark energy (DE) models that have been proposed to explain the data. Instead it is necessary to obtain constraints on the free parameters of
such models and find ways to distinguish between them using observations. One way to test $\Lambda$CDM is to
describe DE as an effective fluid and promote its equation of state
$w$ to a free parameter. This is known as $w$CDM and allows deviations from the standard value of $w=-1$; $w$ is still only measured to
about 5\%-10\% accuracy and has a best-fit value of $w<-1$ for some datasets~\cite{kom11}.

Alternatively, one may go beyond $w$CDM by introducing a new parameter to quantify
interactions within the dark sector. It is natural to expect some new physics in the dark
sector given the richness of
interactions between species in the standard model of particle physics~\cite{pee10}. Indeed taking dark sector interactions to be zero would be an additional assumption of the model. Models of dark sector interactions
(see
e.g.~\cite{wet95,ame99,ame00a,ame00b,hol00,hwa01,zim01,hwa02,toc02,chi03,com03,ame04,far04,koi05,ame06,das06,man06,oli06,sad06,ave07,guo07,bea08,
bea08-1,cor08,oli08,he08,mot08,qua08,que08,abd09,che09,jac09,lav09,li09,xia09,li10,cao10,des10,sim10,bal10b,bal10-1,ame11,avi11,bal11,bal11-1,bal11b,
bal11c,bal11d,ber11,bey11,bey11-1,che11,che11-1,deb11,has11,kos11b,lav11,lee11,lee11-1,li11,li11-1,lip11,mar11,pav11,pel11,pot11,ton11,tar11,sim11,
xu11,zha11,zia11,ave12,cui12}) have little concrete guidance from particle physics, but by studying possible interactions and confronting them
with observations we can shed light on questions such as which models may lead to unphysical behaviour and which are in best agreement with
observations.

Further motivation for interacting dark energy (IDE) models includes: (1)~they may alleviate the coincidence problem of explaining why the domination of DE roughly coincides with the formation of large-scale structure; (2)~IDE affects structure formation and therefore provides a new way to modify the predictions of the standard non-interacting model.

Here we investigate a particular version of the interaction model used
in~\cite{bo08,sch08,val08,ca09a,ca09b,koy09,maj10,val10,bo10}. The general form of this interaction in the background is $\Gamma \bar\rho$, where $\Gamma$ is constant. This has also been used to describe
particle
decays in other contexts~\cite{bil00,mal03,zia04}. Interactions of the form $\alpha H \bar\rho$ (with $\alpha$ constant) (see
e.g.~\cite{he11} and references therein) appear similar, but they mean that the interaction at any event is affected by the global expansion rate $H$, as opposed to the locally determined interaction $\Gamma\bar\rho$.

We describe the IDE model in the background universe in Sec.~\ref{Interactions} and in the perturbed universe in Sec.~\ref{pert}. In
Sec.~\ref{analysis} we investigate the effects of the interaction on the CMB and matter power spectra. We find the best-fit models using CMB, baryon acoustic oscillation (BAO) and supernova (SNIa) data, and  we discuss
the behaviour of some typical models and the implications for the growth of large-scale structure in the Universe. Our conclusions are in Sec.~\ref{conclusions}.

\section{IDE in the background}\label{Interactions}

An IDE model is characterized in the background by the energy transfer rate $\bar Q_x=-\bar Q_c$:
\ba \bar\rho_c'&=&-3\ch\bar\rho_c+a\bar Q_c,\label{rhocprime}
\\\bar\rho_x'&=&-3\ch(1+w)\bar\rho_x+a\bar Q_x,\label{rhoxprime}\ea
where primes denote derivatives with respect to conformal time $\tau$, $\ch=d\ln a/d\tau$ and $w=p_x/\rho_x$. 
We can define an effective DE equation of state parameter to be that of a non-interacting DE with the same $\bar\rho_x(a)$, i.e.
 \be \label{weff}
w_{\rm eff} = w-\frac{a \bar Q_x}{ 3\ch \bar\rho_x}.
 \ee
We can see from Eq (\ref{weff}) that $w_{\rm eff}$ can be dynamical even if $w$ is a constant. Interestingly $w_{\rm eff}$ can be less than $-1$,
or
cross $-1$ during its evolution if $\bar Q_x>0$, even though $w$ itself is always greater than $-1$.

%If $\bar Q_x<0$ then $w_{\rm eff}$ could be $<-1$ even if $w>-1$. 

A simple model for $ \bar Q_x$ is a linear function in the dark sector energy densities. IDE with $\bar Q_x\propto\bar\rho_c$ has been studied in the greatest detail~\cite{bo08,sch08,val08,ca09a,ca09b,koy09,maj10,val10}.
However, for constant $w$ the model suffers from
an instability~\cite{val08}. This instability arises because the model of DE as a fluid with constant $w$ is non-adiabatic. The instability can be cured by allowing $w$ to vary in time~\cite{maj10}.

Here we study the version with $\bar Q_x\propto\bar\rho_x$,
\be \bar Q_x=-\bar Q_c=\Gamma\bar\rho_x,\ee
where $\Gamma$ is a constant transfer rate. The strength of the interaction is measured by $|\Gamma|/H_0$. $\Gamma>0$ corresponds to
energy transfer from DM$\to$DE. This appears somewhat unnatural, since the energy transfer is proportional to $\bar\rho_x$. For $\Gamma<0$, the
interaction can be seen in the background as a decay of DE into DM, which is a more natural model.
The solution of (\ref{rhoxprime}) is~\cite{ca09a}
\be \bar\rho_x=\bar\rho_{x0}a^{-3(1+w)}\exp{[\Gamma(t-t_0)]},\label{rhoxsol}\ee
which shows that $\Gamma>0$ leads to exponential growth of DE. By (\ref{rhocprime}), it follows that $\bar\rho_c$ eventually becomes negative. The
model breaks down if this happens before the current time, which is possible for large $\Gamma/H_0$. Observational constraints require
$\Gamma/H_0\lesssim 1$, so that typically the DM density only becomes negative in the future. In this case, we can treat the model as a viable
approximation, for the past history of the Universe, to some more complicated interaction that avoids the blow-up of DE in the future.
The DE$\to$DM decay model, with $\Gamma<0$, does not have this problem:
both energy densities remain positive at all times when evolving forward from physical initial conditions~\cite{ca09a}.
Furthermore, the $\Gamma<0$ case includes the possibility of beginning with no DM present and having it created entirely from the decay of DE. 

We use a phenomenological fluid model for DE, in which we treat $w$ and the soundspeed $c_s$ as arbitrary parameters. This is a commonly used model
for non-interacting DE, where the model is known as $w$CDM. We impose the condition $w \geq -1$ to avoid `phantom' instabilities that can arise in
scalar field models of DE~\cite{cal02,def10}. The limiting case $w=-1$ is admitted by the background equations, but the perturbation equations have
singularities (see below). Therefore we assume
 \be
w > -1,~~ w=\,\mbox{const}.\label{wgtm1}
 \ee
For completeness, we consider also the $w\leq-1$ case in Appendix \ref{like}.
In the background, the $\Gamma<0$ case appears to be better motivated. However, the analysis of perturbations (see below) shows that these models suffer from an instability when $w>-1$. The $\Gamma>0$ models avoid this instability.

\section{IDE in the perturbed universe}\label{pert}

The critical difference between the background and perturbed IDE is that there is nonzero momentum transfer in the perturbed universe. As emphasized
in \cite{val08}, a model for energy and momentum transfer does not follow from the background model -- and a covariant and gauge-invariant approach is
essential to construct a physically consistent model for energy-momentum transfer.

\subsection*{General IDE}

We give a brief summary of the general discussion in~\cite{val08}.
The perturbed Friedmann metric in a general gauge is
\ba ds^2&=&a^2\big\{-\big(1+2\phi\big)d\tau^2+2\partial_iBd\tau dx^i
\nonumber\\&& +\big[\big(1-2\psi\big) \delta_{ij}+2\partial_i\partial_j E\big]dx^idx^j\big\}.\ea
Each fluid $A$ satisfies an energy-momentum balance equation,
\be \nabla_\nu T^{\mu\nu}_A=Q^\mu_A,\quad \Sigma_AQ^\mu_A=0,\ee
where the second condition expresses conservation of
the total energy-momentum tensor. For dark sector interactions, the energy-momentum transfer four-vectors satisfy
\be Q^\mu_x=-Q^\mu_c .\ee
We split $Q^\mu_A$ relative to the total four-velocity $u^\mu$, so that
\be Q_A^\mu=Q_Au^\mu+F_A^\mu,\quad Q_A=\bar{Q}_A+\delta Q_A,\quad
u_\mu F^\mu_A=0,\label{qadec}\ee
where $Q_A$ is the energy density transfer rate relative to $u^\mu$ and $F_A^\mu$ is the momentum density transfer rate relative to $u^\mu$. To first order
 \be
F_A^\mu=a^{-1}\big(0,\partial^if_A\big),
 \ee
where $f_A$ is the (gauge-invariant) momentum transfer potential.

We choose each $u_A^\mu$ and the total $u^\mu$ as the unique four-velocity with zero momentum density, i.e.
 \ba
&& T^\mu_{A\nu}u^\nu_A=-\rho_Au_A^\mu, ~~
T^\mu_\nu u^\nu=-\rho u^\mu,\\ &&\rho_A=\bar\rho_A+\delta\rho_A, ~~ \rho\equiv \Sigma_A \rho_A=\bar\rho+\delta\rho.
 \ea
Then we have
 \ba \hspace*{-1em}
&& u^\mu_A=a^{-1}\big(1-\phi,\partial^iv_A\big), ~~
u^\mu=a^{-1}\big(1-\phi,\partial^i v\big), \label{4vel}\\ \hspace*{-1em}
&& \big(\Sigma_A\rho_A + \Sigma_A p_A \big)v= \Sigma_A\big(\rho_A+p_A\big)v_A, \label{vva}
 \ea
where $v_A,v$ are the peculiar velocity potentials. Equations (\ref{qadec}) and (\ref{4vel}) imply that
 \ba \hspace*{-2em}
Q_0^A &=&-a\Big[\bar{Q}_A\big(1+\phi\big)+\delta Q_A\Big],\label{zeroth}\\ \hspace*{-2em}
Q_i^A &=& a\partial_i \Big[\bar{Q}_A\big(v+B\big)+f_A\Big]. \label{ith}
 \ea

The evolution equations for
$\delta_A$ and the velocity perturbation
\be
\theta_A=-k^2(v_A+B),
\ee
are \cite{val08}:
\ba \hspace*{-3em}
&&\delta_A'+3\ch(c_{sA}^2-w_A)\delta_A+(1+w_A)\theta_A \nonumber\\ \hspace*{-3em} &&~{}+ 9\ch^2(1+w_A)(c_{sA}^2-c_{aA}^2) \frac{\theta_A}{k^2}
\nonumber\\ \hspace*{-3em} &&~{}-3(1+w_A)\psi'+(1+w_A)k^2(B-E') \nonumber\\ \hspace*{-3em} &&~{}= \frac{a \bar Q_A}{\bar\rho_A}\left[\phi-\delta_A+
3\ch(c_{sA}^2-c_{aA}^2) \frac{\theta_A}{k^2}\right]
+\frac{a}{\bar\rho_A}\delta Q_A, \label{cont}
\\\hspace*{-3em}\nonumber\\ \hspace*{-3em} &&\theta_A'+\ch(1-3c_{sA}^2)\theta_A -\frac{c_{sA}^2}{(1+w_A)} k^2\delta_A
-k^2\phi\nonumber\\ \hspace*{-3em} &&~{}=\frac{a}{(1+w_A)\bar\rho_A} \Big\{ \bar Q_A \big[\theta-(1+c_{sA}^2)\theta_A\big] -k^2f_A \Big\}. \label{eul}
\ea
Here $c_{sA}$ is the physical  soundspeed, defined by $c_{sA}^2= (\delta p_A/ \delta\rho_A)_{\rm rest frame}$, and $c_{aA}$ is the adiabatic
soundspeed, defined by $c_{aA}^2\equiv \bar p_A'/\bar\rho_A'$. For the adiabatic DM fluid, $c_{sc}^2=c_{ac}^2=w_c=0 $. By contrast, the DE fluid is
non-adiabatic: $c_{ax}^2=w<0$ and so $c_{ax}$ cannot be the physical soundspeed. The physical soundspeed for the fluid DE model is a phenomenological
parameter. It must be real and non-negative to avoid unphysical instabilities. We choose $c_{sx}=1$, which is the sounspeed for quintessence (a
self-consistent model of DE). Our analysis is insensitive to the value of $c_{sx}$, as long as $c_{sx}$ is close to one, so that DE does not cluster
significantly on sub-Hubble scales. (See~\cite{val08} for more details.)

\subsection*{DM-baryon bias from IDE}

In IDE models, the DE exerts a drag on DM but not on baryons. This leads to a linear DM-baryon bias in the late-time density perturbations, and in
general also to a velocity difference \cite{koy09}. For baryons after decoupling
 \be
\delta_b'+\theta_b-3\psi'+k^2(B-E')=0,~ \theta_b'+\ch\theta_b-k^2\phi=0.
 \ee
Thus for non-interacting DE models, 
 \be
\theta_c-\theta_b= (\theta_c-\theta_b)_{\rm{i}} \frac{a_{\rm{i}}}{a}
 \ee
We can choose $(\theta_b-\theta_c)_{\rm{i}}=0$, so that
 \be
\theta_c-\theta_b=0,~~~ \delta_c-\delta_b=(\delta_c-\delta_b)_{\rm{i}}.
\ee
Thus in standard DE models, there is no DM-baryon velocity difference, and any linear density perturbation difference is determined by initial conditions.

For IDE models, the interaction induces a non-constant difference between $\delta_c$ and $\delta_b$ -- which is degenerate with the standard galaxy
bias. The Euler equation for DM is (\ref{eul}), with $c_{sc}^2=w_c=0$. This differs from the standard Euler equation unless
$k^2f_c=\bar{Q}(\theta-\theta_c)$, which follows only for $Q^\mu_c=Q_cu^\mu_c$, regardless of the form of $Q_c$~\cite{koy09}. In those models that
modify the Euler equation for DM, there will also be a velocity bias. Equations (\ref{cont}) and (\ref{eul}) imply
 \ba
&&(\delta_c-\delta_b)'+ \frac{a\bar Q_c}{\bar\rho_c} (\delta_c-\delta_b) +(\theta_c-\theta_b)  \nonumber\\
&&~~~~~~{}= \frac{a}{\bar\rho_c}\Big[\delta Q_c+ \bar Q_c(\phi- \delta_b) \Big], \label{dmbd} \\
&& (\theta_c-\theta_b)'+ \Big(\ch+ \frac{a\bar Q_c}{\bar\rho_c}\Big) (\theta_c-\theta_b)  \nonumber\\
&&~~~~~~{}= \frac{a}{\bar\rho_c}\Big[-k^2 f_c+ \bar Q_c(\theta- \theta_b) \Big]. \label{dmbv}
 \ea
Thus there will be a velocity bias, unless $Q^\mu_c=Q_cu^\mu_c$.

\subsection*{Our IDE models}

The preceding equations are completely general.
A choice must now be made for the energy-momentum transfer in the dark sector. Firstly, the nature of the background energy transfer suggests that we take
 \be
Q_x=\Gamma\rho_x= \Gamma\bar\rho_x(1+\delta_x)=-Q_c,
 \ee
where $\delta_A \equiv \delta\rho_A/\bar\rho_A$. Thus we are treating $\Gamma$ as a universal constant.
For the momentum transfer,
the simplest physical choice is that there is no momentum transfer in the rest frame of either DM or DE~\cite{val08,koy09}. This leads to two types of model, with energy-momentum transfer four-vectors parallel to either the DM or the DE four-velocity:
 \ba
Q^\mu_x&=& Q_x u^\mu_c=-Q^\mu_c ~~~\mbox{type:  \qc}, \label{qc}
 \\
Q^\mu_x &=& Q_x u^\mu_x=-Q^\mu_c ~~~\mbox{type:  \qx}. \label{qx}
 \ea
Thus
 \be Q_\mu^x=a\Gamma\bar\rho_x\big[ 1+\delta_x+\phi,\,\partial_i(v_A+B)\big],\ee
where $A=c,x$  for type \qc, \qx. By (\ref{ith}), the momentum transfer relative to the background frame is
 \ba \label{fxc}
f_x &=& \Gamma\bar\rho_x(v_c-v)=-f_c ~~~\mbox{for \qc},\\ 
f_x &=& \Gamma\bar\rho_x(v_x-v)=-f_c ~~~\mbox{for \qx}. \label{fxx}
 \ea

For both the \qc and \qx models, the density perturbation (continuity) equation (\ref{cont}) reduces to
\ba \hspace*{-3em} && \delta_c'+\theta_c-3\psi'+k^2(B-E')= a\Gamma\frac{\bar\rho_x}{\bar\rho_c} (\delta_c-\delta_x-\phi), \label{deltacprime} \\
\label{deltaxprime} \hspace*{-3em}
&& \delta_x'+3\ch(1-w)\delta_x+(1+w)\theta_x
+9\ch^2(1-w^2)\frac{\theta_x}{k^2}
\nonumber\\ \hspace*{-3em} &&~{}-3(1+w)\psi'+(1+w)k^2(B-E')\nonumber \\ \hspace*{-3em} &&~{} =a\Gamma\Big[\phi +3\ch(1-w)\frac{\theta_x}{k^2}\Big].
\ea
The velocity perturbation (Euler) equations are however different. For the \qc model, (\ref{eul}) gives
\ba \hspace*{-3em} && \theta_c'+\ch\theta_c-k^2\phi=0,\label{thetacprimec}\\
\hspace*{-3em} &&\theta_x'-2\ch\theta_x-\frac{k^2\delta_x}{(1+w)} -k^2\phi
=\frac{a\Gamma}{(1+w)} \big(\theta_c-2\theta_x \big). \label{thetaxprimec} \ea
For the \qx model:
\ba \theta_c'+\ch\theta_c-k^2\phi &=& a\Gamma\frac{\bar\rho_x} {\bar\rho_c} (\theta_c-\theta_x),\label{thetacprimex}\\
\label{thetaxprimex}\theta_x'-2\ch\theta_x-\frac{k^2\delta_x} {(1+w)}-k^2\phi &=& -\frac{a\Gamma\theta_x}{(1+w)}.\ea
It follows that the Euler equation for DM in the \qc model has the standard form, whereas it is modified in the \qx model.

\subsection*{Instability}

There is an obvious instability in
the Euler equations for DE, (\ref{thetaxprimec}) and (\ref{thetaxprimex}), as $w \to-1$. Thus we must exclude the value $w=-1$. This instability is
different from that for a dynamical DE model with $w$ crossing $-1$, in which case the DE perturbation is well-defined, but at least one more degree
of
freedom is required, usually leading to its interpretation as a sign of modified gravity. Here though, the DE is not dynamical and the DE
perturbation is ill-defined at $w=-1$.

These equations also reveal an instability for $w\neq-1$ in certain regions of parameter space. The underlying cause of this instability is the choice
of $c_{sx}^2=1$, which means that the DE fluid is non-adiabatic, as discussed above. It is qualitatively similar to the
instability first discovered for constant $w$ IDE in~\cite{val08}. (See also~\cite{gav09,he09,jac09,lop09,gav10,lop10a} for the case of  models with
$\Gamma$ replaced by $\alpha H$). This is a DE velocity instability, which then drives an instability in the DE and DM density perturbations.

On large scales,  we can drop the $\delta_x$ and $\phi$ terms in the DE Euler equations (\ref{thetaxprimec}) and (\ref{thetaxprimex}).  In
(\ref{thetaxprimec}) we can also set $\theta_c=0$ by (\ref{thetacprimec}). Then we can integrate to find that
 \be \label{instab}
\frac{\theta_x}{\theta^{(\Gamma=0)}_x} =  \exp\left[ - \alpha\frac{\Gamma}{1+w} (t-t_0)\right],
 \ee
where $\theta^{(\Gamma=0)}_x$ is the DE velocity in the non-interacting case, and $\alpha=2,1$ for \qc, \qx.  It follows that
 \be
-\frac\Gamma{(1+w)}>0~~\Rightarrow ~~ \mbox{instability}.
 \ee
Note that although one can choose a reference frame where $\theta_x \equiv 0$, the instability is still present in the velocity difference, which is gauge invariant. Given our assumption (\ref{wgtm1}), the stable models must have positive $\Gamma$, i.e.
 \be
w>-1~\mbox{and}~\Gamma > 0 ~~\Rightarrow ~~ \mbox{no instability}, \label{noinst}
 \ee
for both \qc and \qx. This defines for us the physically acceptable models. 
In Appendix \ref{like} we allow for any sign of $\Gamma$ and $1+w$. In order for the
instability to affect significantly the perturbation evolution by
today, the time scale of growth of $\theta_x$ in (\ref{instab}) should be shorter
than the Hubble time, i.e., the models with
 \be
-\frac{\Gamma}{H_0(1+w)} \gtrsim \Bigg\{
\begin{array}{cl}1 & \mbox{for \qc}, \\& \\ 2 & \mbox{for \qx}. \end{array} \label{instab2}
 \ee
may not be viable. The results from our full parameter scan confirm
this (see the excluded wedges near to $w=-1$ in Fig. \ref{3d}).

\section{Analysis}\label{analysis}

The evolution of $\Gamma w$CDM models was computed numerically using a modified version of the CAMB Boltzmann code \cite{lew00}, including implementation of the
initial conditions derived in Appendix \ref{ini}. The code was adapted (a)~to allow for the
non-standard background evolution caused by the interactions; (b)~to evolve the DM velocity perturbation (ordinarily set to zero); (c) to
suppress perturbations when $|1+w|<0.01$ due to the blow up of terms in~(\ref{thetaxprimec}) and (\ref{thetaxprimex}) as
$w\rightarrow-1$. It is useful to include the $w=-1$ limit for comparison with $\Lambda$CDM. 
\begin{table*}
\begin{center}
\begin{tabular}[c]{c|c|c|c|c|c|c|c|c|c|l}
Model &$Q_A^\mu$&$\Delta\chi^2$&$\Gamma/H_0$&$w$&$H_0$&$\Omega_bh^2$& $\Omega_ch^2$&$n_s$&$A_s$&$\tau_{rei}$\\\hline
$\Lambda$CDM best-fit&-&0&-&$-$1&69.8&0.0223&0.113&0.960&2.16$\times10^{-9}$&0.0844\\\hline
$\Lambda$CDM69&-&0.774&-&$-$1&69.0&0.0221&0.114&0.958& 2.18$\times10^{-9}$&0.0855\\\hline
$\Lambda$CDM70&-&$-$0.0200&-&$-$1&70.0&0.0224&0.112&0.962& 2.16$\times10^{-9}$&0.0844\\\hline
$w$CDM best-fit&-&$-$0.220&-&$-$1.03&70.7&0.0222&0.113&0.960&2.18$\times10^{-9}$&0.0883\\\hline
$\Gamma w$CDM A& \qc &-&0&$-$0.98&70.0&0.0226&0.112&0.960&2.10$\times10^{-9}$&0.0900\\\hline
$\Gamma w$CDM B& \qc &-&0.2&$-$0.98&70.0&0.0226&0.112&0.960&2.10$\times10^{-9}$&0.0900\\\hline
$\Gamma w$CDM C& \qc &-&0.4&$-$0.98&70.0&0.0226&0.112&0.960&2.10$\times10^{-9}$&0.0900\\\hline
$\Gamma w$CDM 1a& \qc &$-$0.00830&0.4&$-$0.95&70.9&0.0222&0.0702&0.961&2.16$\times10^{-9}$&0.0816\\\hline
$\Gamma w$CDM 1b& \qc &0.702&0.7&$-$0.85&70.0&0.0223&0.0311&0.963&2.15$\times10^{-9}$&0.0832\\\hline
$\Gamma w$CDM 2a& \qx &$-$0.236&0.4&$-$0.95&71.0&0.0224&0.0701&0.966&2.19$\times10^{-9}$&0.0870\\\hline
$\Gamma w$CDM 2b& \qx &$-$0.0420&0.7&$-$0.85&70.2&0.0224&0.0305&0.966&2.15$\times10^{-9}$&0.0819\\\hline
%$\Gamma w$CDM 3& \qx %&0.0879&-0.5&-1.2&71.4&0.0221&0.159&0.955&2.17$\times10^{-9}$&0.0827\\\hline
$\Gamma\ge0, w\ge-1$ best-fit& \qc &$-$0.0522&0.366&$-$0.964&71.0&0.0224&0.0748&0.963&2.18$\times10^{-9}$&0.0849\\\hline
$\Gamma\ge0, w\ge-1$ best-fit& \qx &$-$0.322&0.798&$-$0.851&70.4&0.0224&0.0194&0.965&2.18$\times10^{-9}$&0.0870\\
%$\Gamma\ge0, w\ge-1$ best-fit& \qc &-0.0420&0.206&-0.974&70.8&0.0224&0.0907&0.964&2.17$\times10^{-9}$&0.0867\\\hline
%$\Gamma\ge0, w\ge-1$ best-fit& \qx &-0.314&0.793&-0.850&70.6&0.0225&0.0193&0.966&2.17$\times10^{-9}$&0.0842\\
%$\Gamma\ge0, w\ge-1$ mean& \qc &-&0.493&-0.909&70.3&0.0224&0.0578&0.964&2.17$\times10^{-9}$&0.0858\\\hline
%$\Gamma\ge0, w\ge-1$ mean& \qx &-&0.533&-0.905&70.3&0.0223&0.0533&0.963&2.18$\times10^{-9}$&0.0862\\

\end{tabular}
\end{center}
\caption{Cosmological parameters for IDE models (see Appendix~\ref{like} for more general
constraints).\label{models}}
\end{table*}

\begin{figure*}
 \includegraphics[width=\columnwidth]{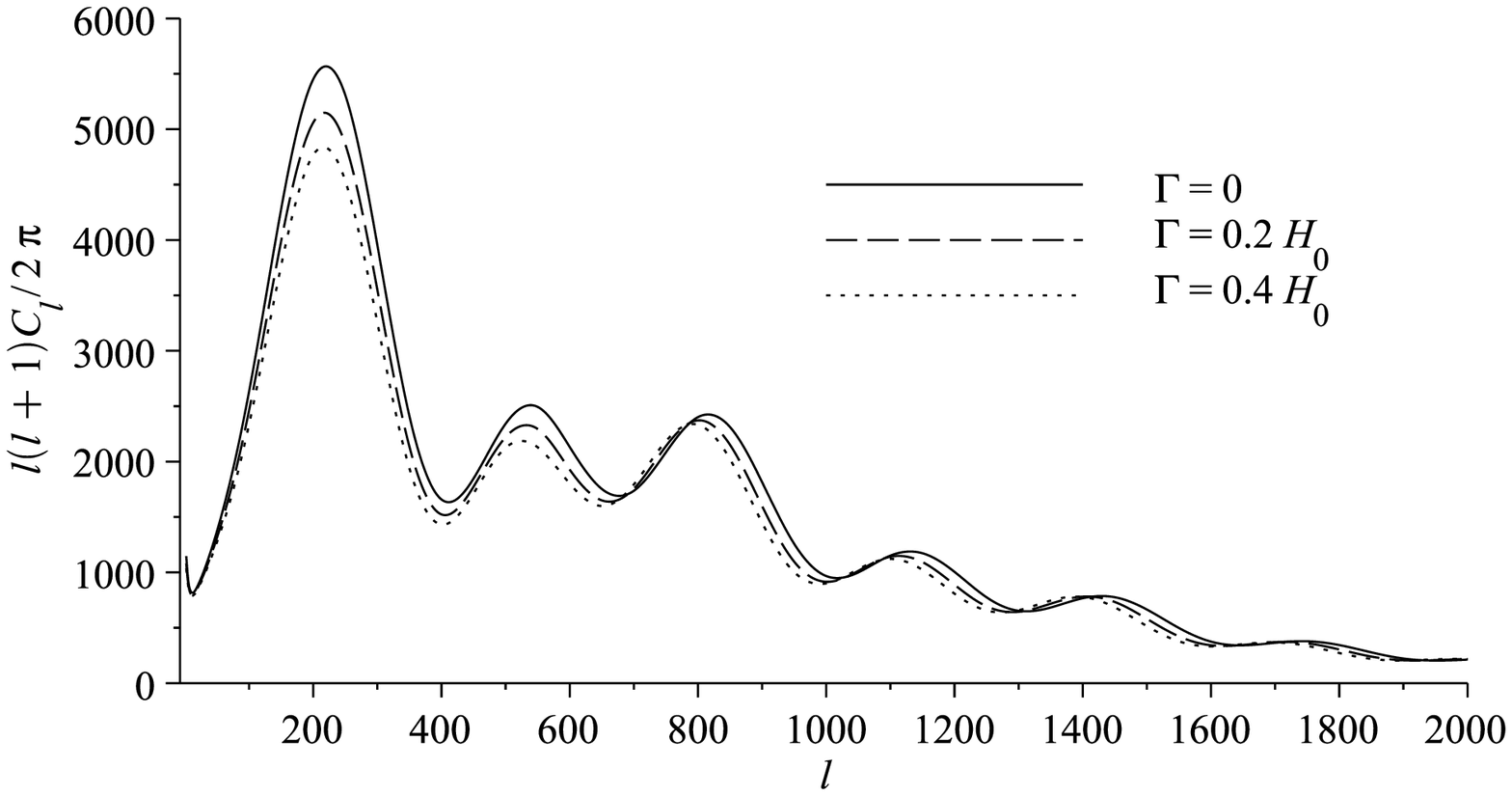}
 \includegraphics[width=\columnwidth]{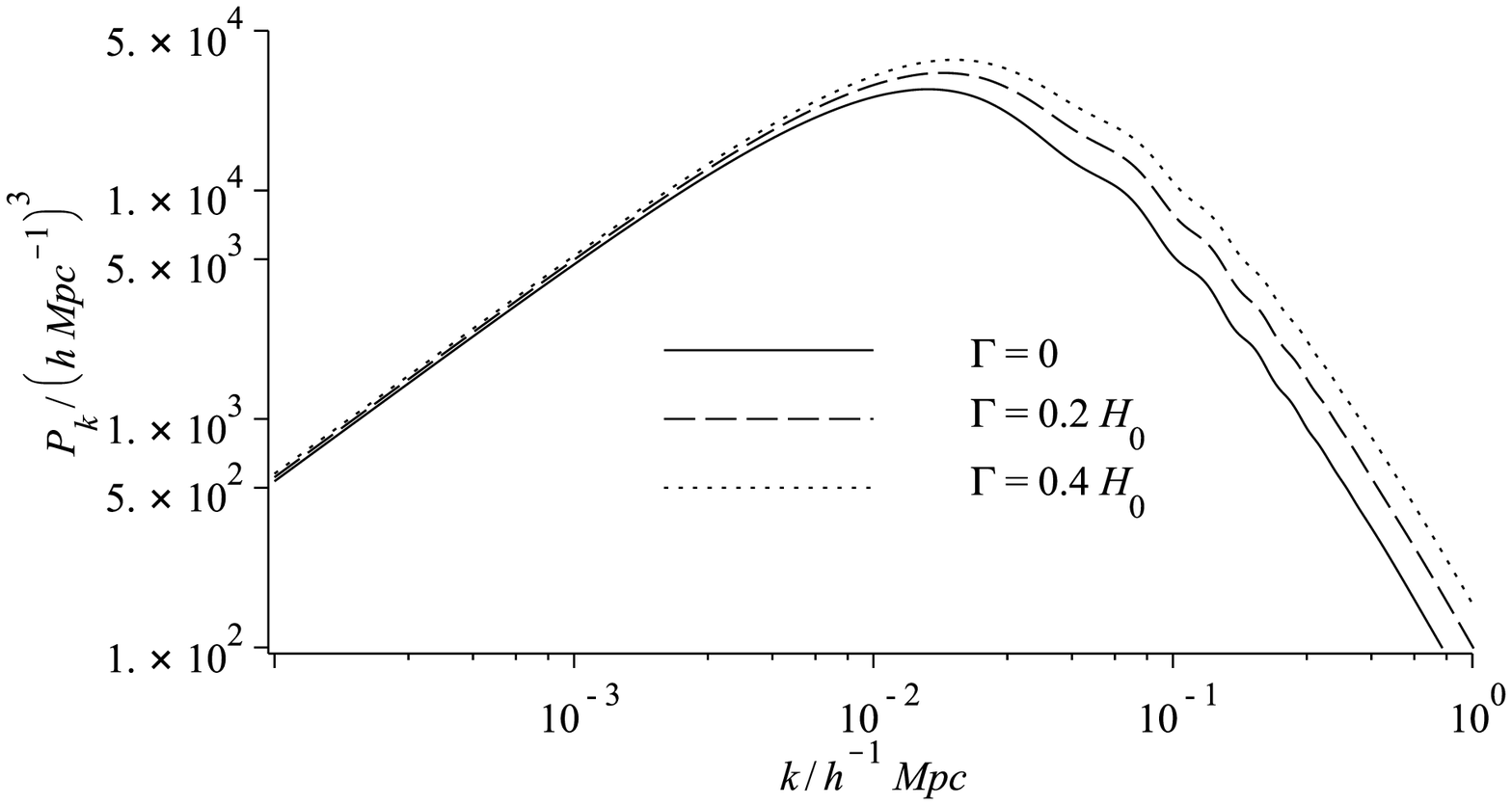}
 \caption{CMB and total matter power spectra from the modified CAMB code for 3 \qc models with different values of $\Gamma$ but identical values of
their remaining parameters (see $\Gamma w$CDM A,B,C in Table~\ref{models}). \label{comparison}}
\end{figure*}
\begin{figure*}
 \includegraphics[width=\columnwidth]{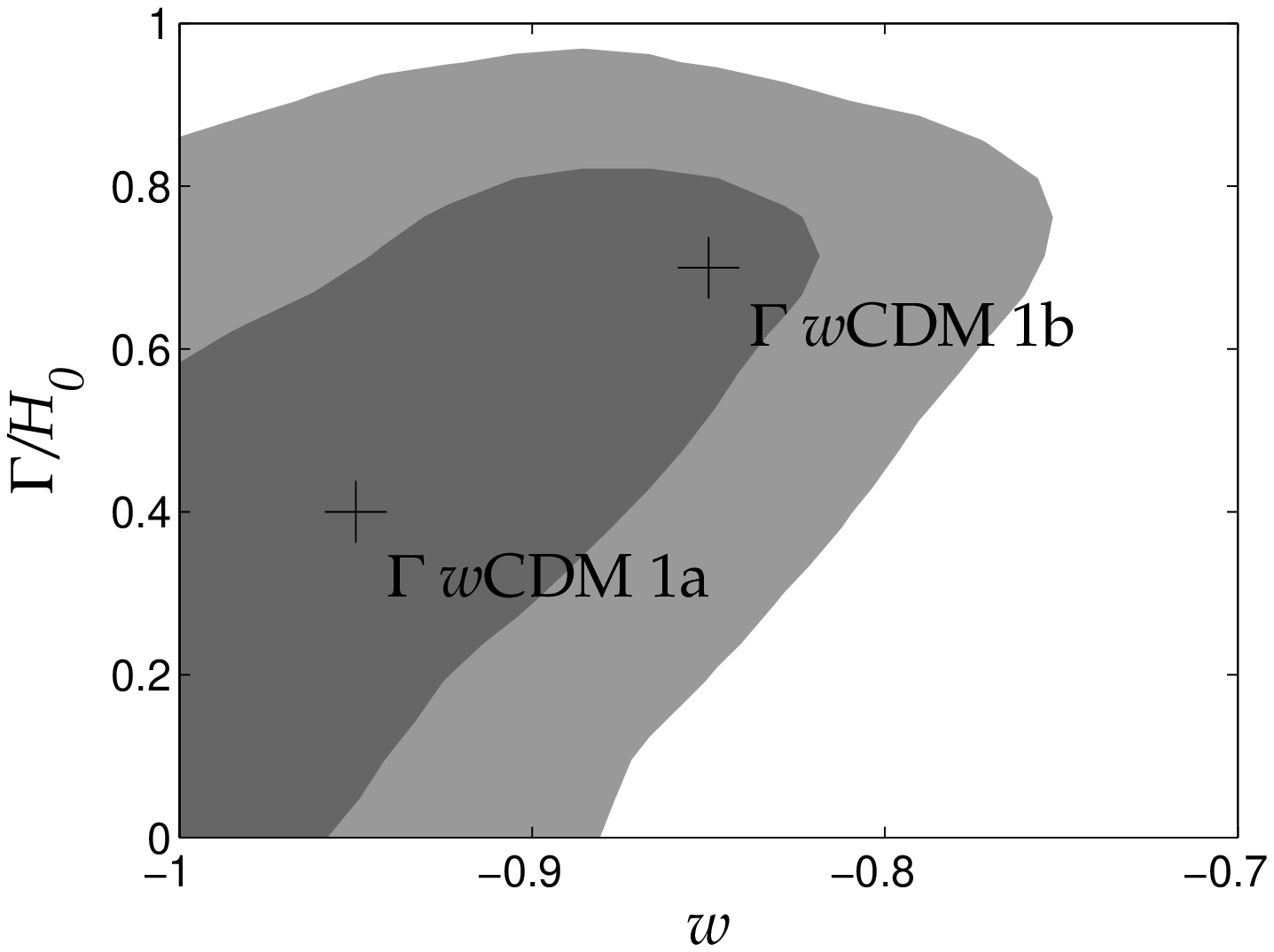}
 \includegraphics[width=\columnwidth]{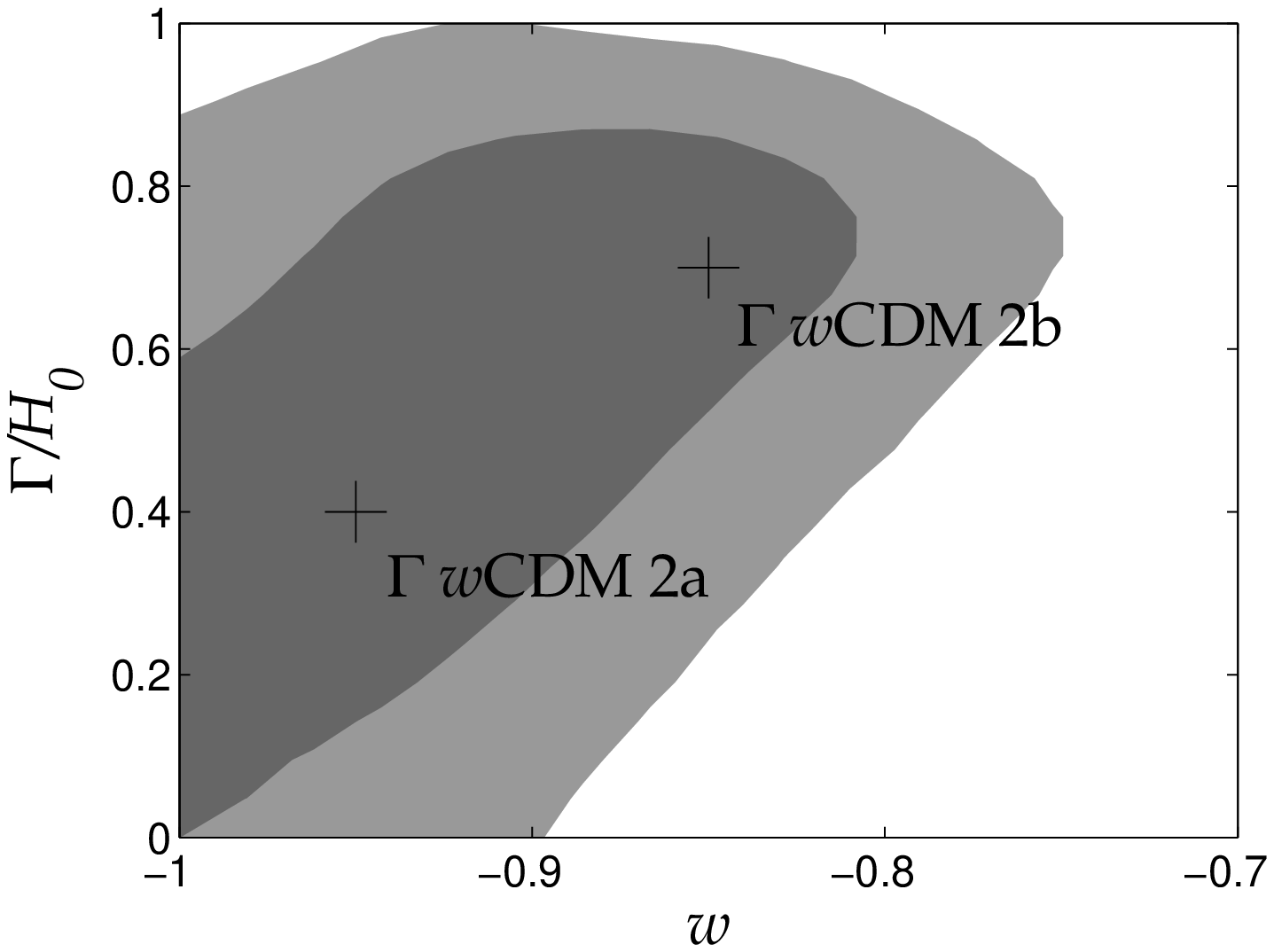}
\caption{Smoothed 68\% and 95\% contours of the marginalised probability distribution for IDE model with \qc (left) and  \qx (right)
in the range of stability, $w>-1$ and $\Gamma\ge0$. Crosses identify models chosen to be analyzed in more detail (see Table~I). \label{likex}}
 \end{figure*}
 \begin{figure*}
 \includegraphics[width=\columnwidth]{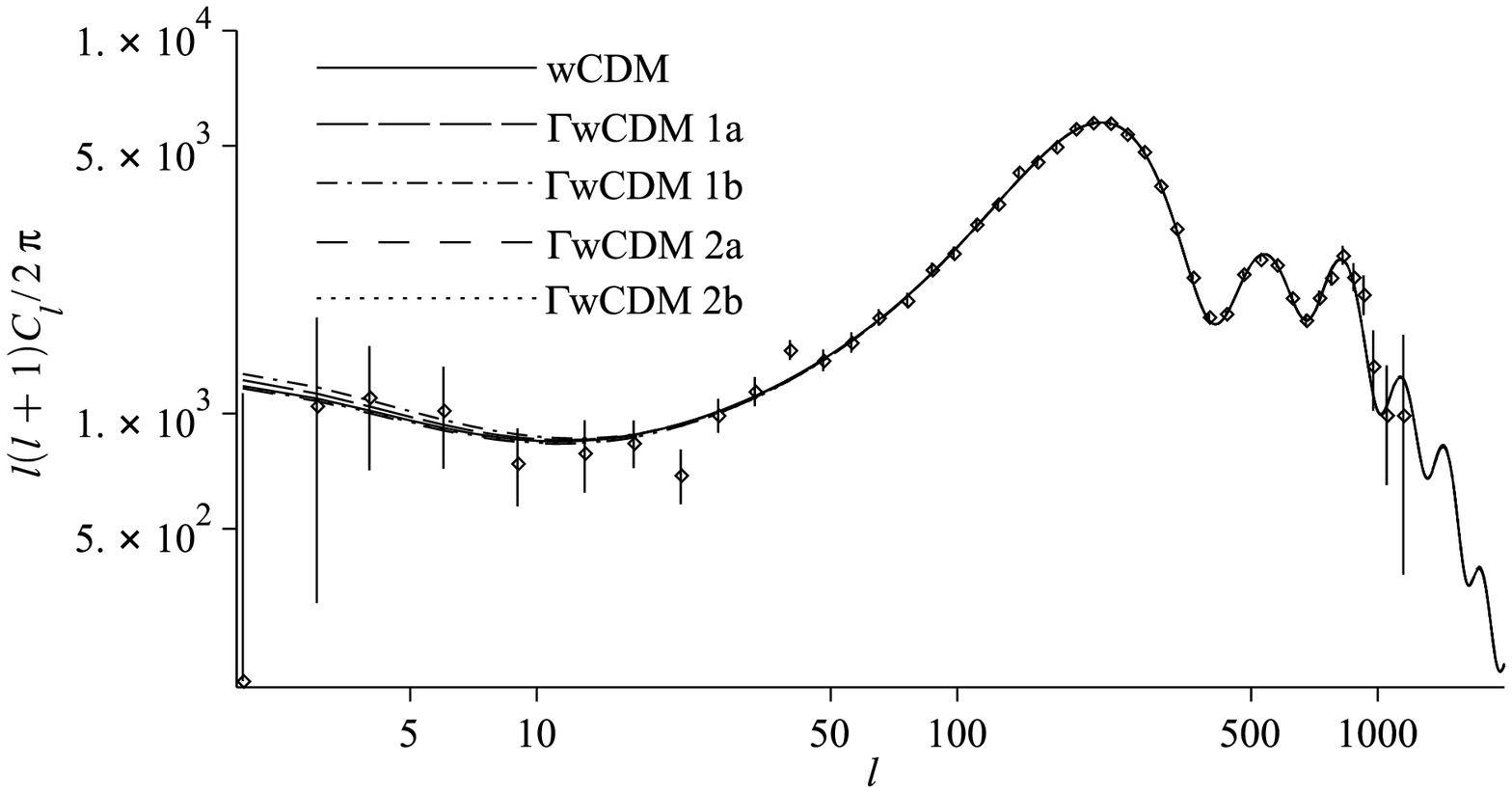}
 \includegraphics[width=\columnwidth]{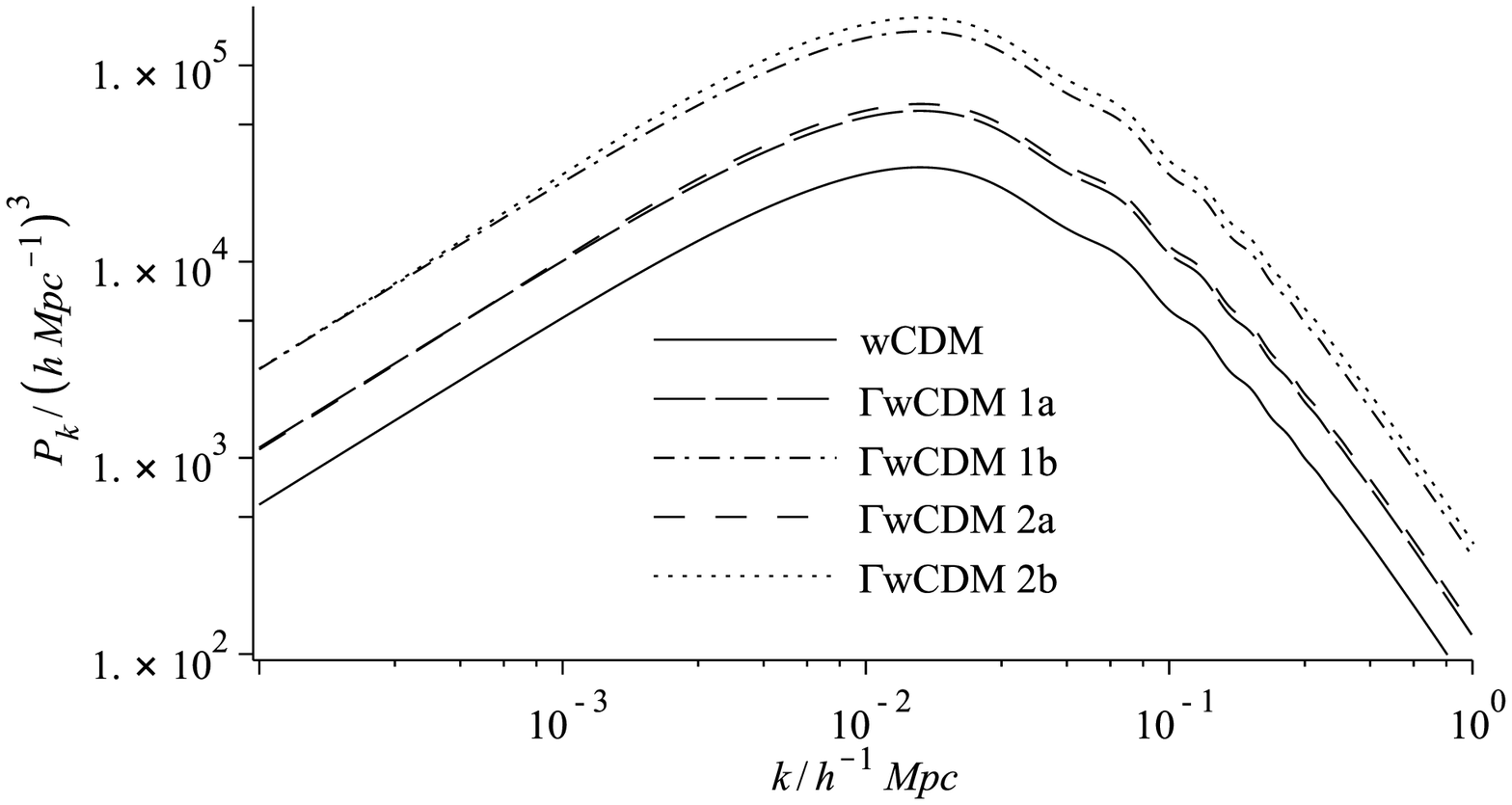}%
\caption{CMB and total matter power spectra from the modified CAMB code for the WMAP7 $w$CDM best-fit values and the $\Gamma w$CDM 1a,1b,2a,2b models
chosen from the 95\% confidence range for further analysis (see Table~\ref{models}). The best-fit values of standard cosmological parameters
were found using CosmoMC. Models 1a,1b have $\Gamma=0.4H_0$ and \qc while 2a,2b have $\Gamma=0.7H_0$ and \qx.\label{best}}
 \end{figure*}
 \begin{figure*}
 \includegraphics[width=0.8\columnwidth]{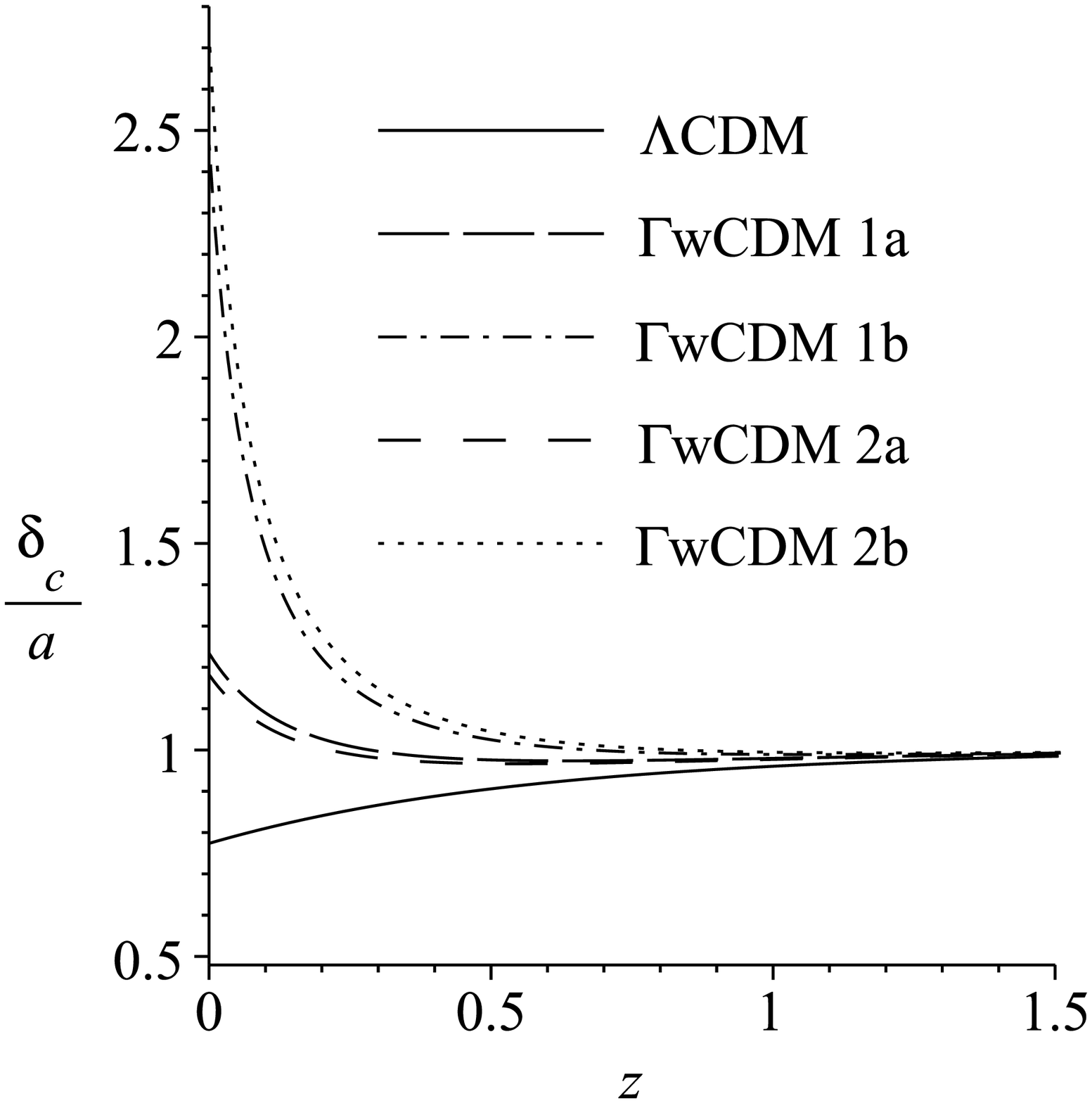}\quad\quad\quad \quad\quad\quad
 \includegraphics[width=0.8\columnwidth]{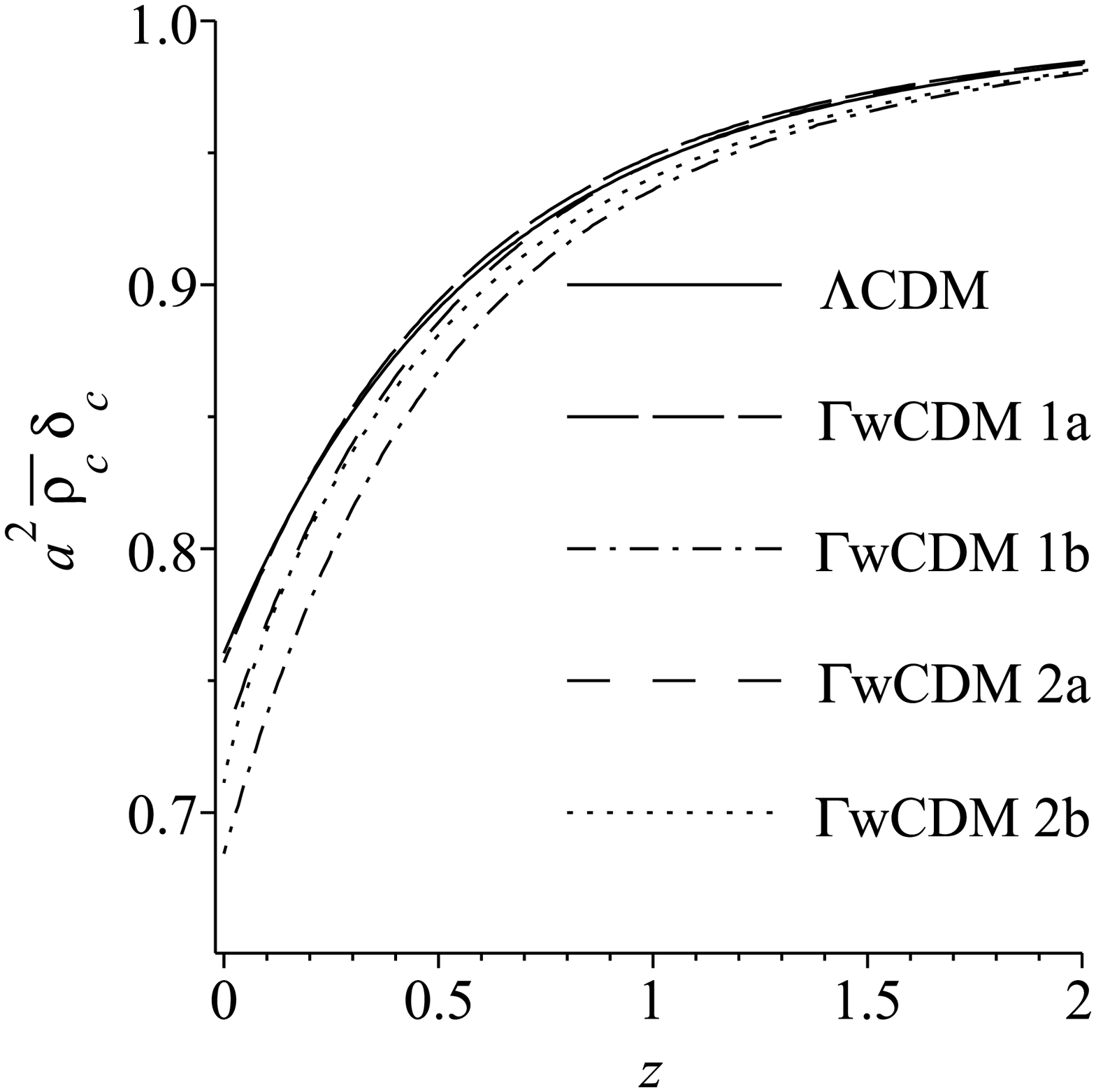}
 \caption{{\em Left:} Normalized growth rates for $\Lambda$CDM and the same best-fit models as in Fig. \ref{best}.  {\em Right:} The same models but showing a normalized combination of
$a^2\bar{\rho_c}\delta_c$ which is important for the ISW effect. \label{a2dcrc}}
 \end{figure*}
  \begin{figure*}
 \includegraphics[width=0.8\columnwidth]{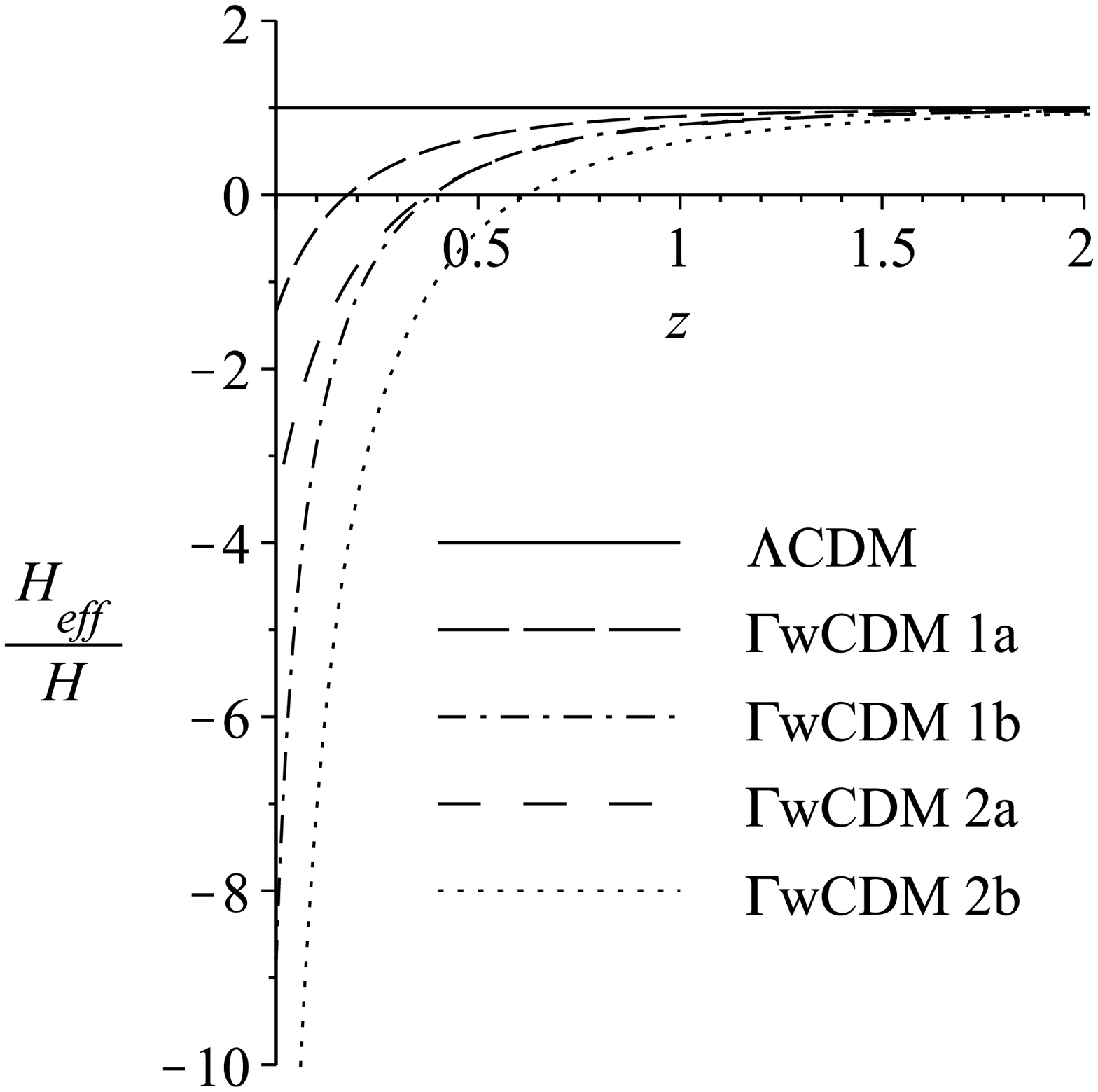}\quad\quad\quad \quad\quad\quad
 \includegraphics[width=0.8\columnwidth]{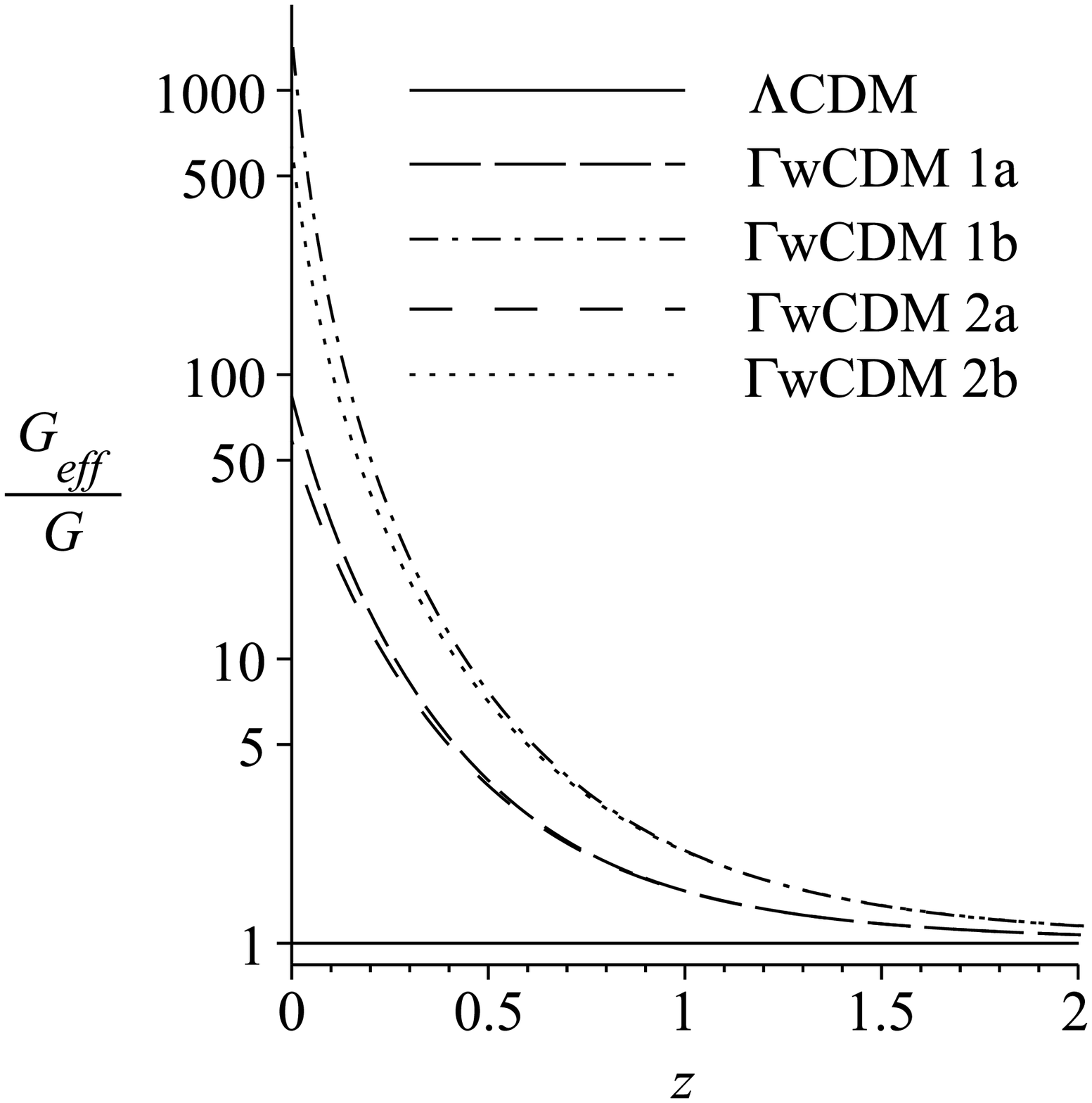}
 \caption{Deviations from $\Lambda$CDM of the effective Hubble parameter (left) and effective Newton constant for $\delta_c$ (right), for the same best-fit models as in Fig. \ref{best}.  \label{effs}}
 \end{figure*}
  \begin{figure*}
 \includegraphics[width=0.8\columnwidth]{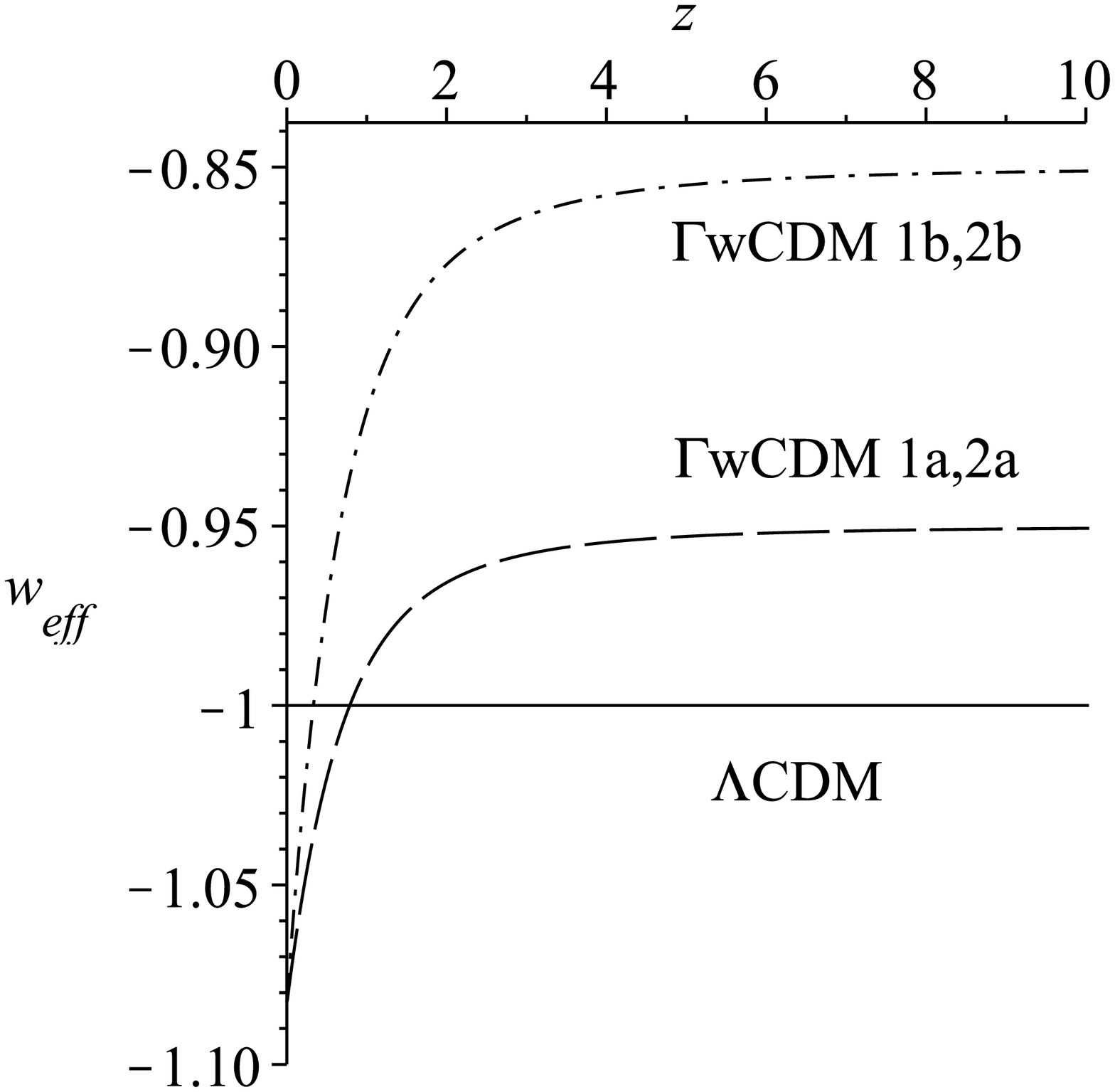}\quad\quad\quad \quad\quad\quad
 \includegraphics[width=0.8\columnwidth]{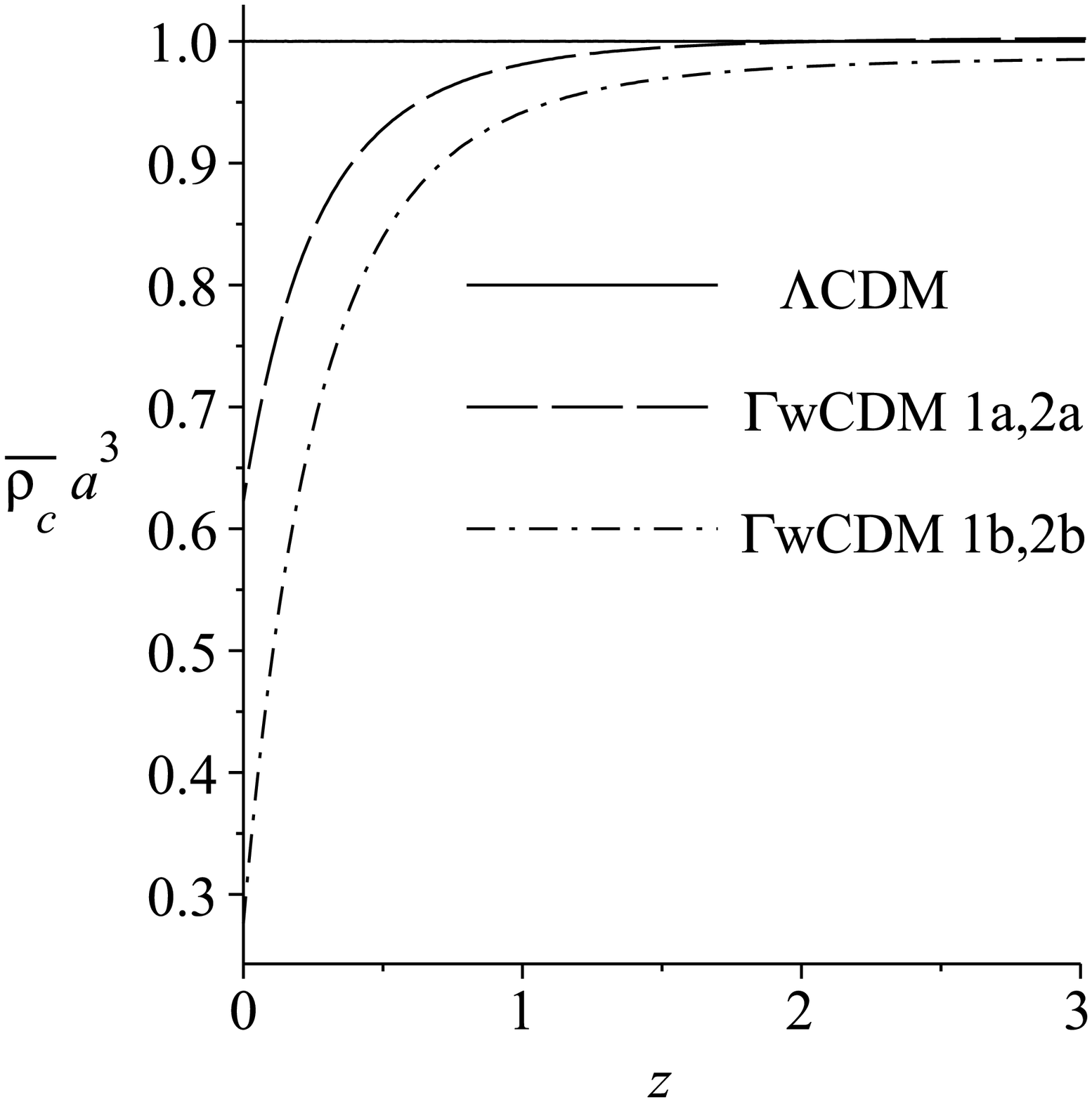}
 \caption{Comparison with $\Lambda$CDM of the effective DE equation of state (left) and the DM density (right), for the same best-fit models as in Fig. \ref{best}.  \label{effs2}}
 \end{figure*}
Insight into the physical implications of the interaction can be gained by running the modified CAMB code with fixed input parameters, varying only
the interaction rate $\Gamma$. Figure~\ref{comparison} shows the CMB power spectrum for three values of $\Gamma$ with all other cosmological parameters
set to typical values (see Table~\ref{models} for details).

Positive $\Gamma$ describes a transfer of energy from DM to DE, so with fixed $\Omega_c$
today, the DM energy density
would have been correspondingly greater in the past than without interactions. Hence the amplitude of the CMB power spectrum is decreased and the position of the peaks shifted, since a larger proportion of DM at early times implies a smaller amount of baryonic matter and therefore a more significant effect from photon
driving before decoupling. The present-day matter power spectrum for these choices of $\Gamma$ shows that a relative increase in the past DM density naturally leads to more structure formation and an increase in the amplitude of the matter power spectrum.

The modified CAMB code was integrated into the CosmoMC~\cite{lew02} Markov-Chain Monte-Carlo (MCMC) code in order to explore the
parameter space (see Appendix \ref{like} for more details). The data used in the MCMC analysis were: CMB (WMAP7~\cite{kom11}), BAO~\cite{per07}, HST~\cite{rie09},
and SNIa (SDSS~\cite{kes09}) data, as well as a prior on $\Omega_b$ from big-bang nucleosynthesis~\cite{bur01}. Figure~\ref{likex}
shows the 68\% and 95\% likelihood contours in the $w-\Gamma$ plane for the two different momentum transfer models \qc and \qx, where all other parameters have been marginalized over.

The likelihood regions are very similar for the two models since they differ only in their perturbations and the observations predominantly constrain
the background evolution. The best-fit values for the \qc and \qx models are different due to the
Integrated Sachs-Wolfe (ISW) effect on the CMB, as shown in Table~\ref{models}. For the \qx model the best-fit value is a genuine global maximum. For
the \qc model however the mean likelihood function of $w$ and $\Gamma$ is essentially one-tailed, with the true global maximum lying outside of the
region we consider physical (see Appendix~\ref{like}). Indeed the $\chi^2$ of this point is close to that of $\Lambda$CDM (see Table~I).
This is because \qc models in this region can closely mimic the ISW signature of $\Lambda$CDM.

The CMB data is best fit by a particular ISW signal,
and the two momentum transfer models differ somewhat in their structure formation histories. DM in the \qx model receives a change in momentum from
the DE
perturbations, as expressed by its modified Euler equation (\ref{thetacprimec}), leading to more structure growth relative to the \qc
model. This means that DE can be weaker for the \qx model in order to give the same amount of ISW signal as the \qc model.

In order to assess the relative merits of these models we have included the change in $\chi^2$ from a $\Lambda$CDM baseline. To
help put this quantity into context we have also included two best-fit $\Lambda$CDM models with $H_0$ fixed at 69 and 70~km/s/Mpc.
The mean likelihoods of the samples vary little in the direction of the degeneracy in the $w-\Gamma$ plane. For example, the difference in
$\Delta\chi^2$ between the \qx best-fit and the $\Lambda$CDM best-fit is less than
the difference between the two fixed-$H_0$ $\Lambda$CDM models ($\Lambda$CDM69 and $\Lambda$CDM70 in Table~\ref{models}).

In Fig. 6, we show the effective DE equation of state and the $a^3$-scaled energy density for DM for a selection of models in comparison with
$\Lambda$CDM. Interestingly, we find that the $w_{\rm eff}$ for the best-fit $\Gamma w$CDM models with $w=-0.85$ and $-0.95$ crosses $-1$ during its
evolution, showing a quintom-like behaviour~\cite{fen05}.

We have focused on the stable $\Gamma>0$ models with $w>-1$. These models do have
a problem of negative DM energy densities in the future, but we assume that this can be cured by a more realistic model to which our model is a good approximation when $\rho_c>0$.

\subsection*{Growth of structure}\label{best-fit}

Models $\Gamma w$CDM 1a,1b,2a,2b (see Table~\ref{models}) were selected for further study. CosmoMC was rerun with $\Gamma$ and $w$ fixed, to obtain the best-fit values of the other non-derived
parameters for input back into CAMB, namely $\Omega_bh^2$, $\Omega_ch^2$, $H_0$, $n_s$ (scalar spectral index), $A_s$ (scalar amplitude) and $\tau_{\rm rei}$  (optical depth of reionization).

Figure~\ref{best} shows the CMB power spectrum of the best-fit parameter sets for the chosen values of $\Gamma$ and $w$. The only significant
difference between the CMB spectra is in the ISW feature, although this is not very large because CosmoMC has fit them well to the data from
WMAP7. By contrast, there are dramatic differences between the total matter power spectra at $z=0$ for these models (see also Fig. \ref{a2dcrc}). We chose not to fit the matter power
spectrum to observational data -- because the modification to the
growth of matter perturbations $\delta_m$ due to the interactions is degenerate with the galaxy-DM bias $b$ in observations of galaxy number
density fluctuations: $\delta_g=b\delta_m$. This degeneracy is governed by equations (\ref{dmbd}) and (\ref{dmbv}). Figure~\ref{best} does not include any bias.

Note that $\Omega_{c}$
can be very small in models with large $\Gamma$, since it can be compensated for by a higher $w$ in order to obtain a sensible $H_0$. This explains the correlation
in the $\Gamma-w$ plane shown in Fig.~\ref{likex}, so the late-time effect of the DM may be proportionately
even greater than one might think at first glance.

The combination of similar ISW signatures and large differences in the growth of structure is unusual -- in a $\Lambda$CDM cosmology for
example, different growth rates lead to correspondingly dissimilar ISW signatures. The mechanisms behind this are clearest from the growth of DM perturbations in the Newtonian limit: on sub-Hubble scales at late-times,
\be \delta_c\gg\phi=\psi,~~\delta_x=\phi'=\psi'=0,\ee
in the Newtonian gauge ($B=0=E$). The evolution of synchronous gauge density perturbations in CAMB matches that of perturbations in the Newtonian
gauge. The ISW effect comes from gravitational potentials determined by the Poisson equation,
\be k^2\phi=-4\pi G a^2 (\rho_c\delta_c+\rho_b\delta_b), \label{poisson}\ee
and the left panel of Fig.~\ref{a2dcrc} shows that there are indeed large differences between the models in their growth rates at late times. The reason the ISW
effects can remain small for these models is that the non-standard
background evolution (see Fig. \ref{effs2}) can counteract the growth of $\delta_c$ in (\ref{poisson}) and lead to relatively stable gravitational
potentials. The right panel of Fig.~\ref{a2dcrc} shows that
the relevant combination, $a^2\bar\rho_c\delta_c$, can remain comparable for models with very different structure formation histories such as those
considered here. Note how well the $\Gamma w$CDM 1a model mimics the $\Lambda$CDM behaviour of $a^2\bar\rho_c\delta_c$, effectively leading to the
same $\chi^2$ (see Table~I).

This important feature of IDE models has implications for any cosmological test which assumes a standard evolution of the DM energy density during matter domination, such as those for detecting
deviations from GR. It may also be useful for distinguishing between
IDE and modified gravity models~\cite{tsu10}, which have standard background evolutions.

Using (\ref{rhocprime}), (\ref{rhoxprime}), (\ref{deltacprime}), (\ref{deltaxprime}), (\ref{thetacprimec}), (\ref{poisson}) and the Friedmann equation, $\ch^2=8\pi Ga^2\rho_{\rm tot}/3$,
a velocity independent equation of motion for $\delta_c$ can be derived for the \qc model:
\ba &&\delta_c''+\ch\Big(1-\frac{a\Gamma}{\ch}\frac{\rho_x}{\rho_c} \Big)\delta_c'= 4\pi Ga^2\Big\{\rho_b\delta_b +\rho_c\delta_c\Big[1+
\nonumber \\ &&~{}
\frac{2\rho_{\rm tot}}{3\rho_c} \frac{a\Gamma}{\ch}\frac{\rho_x}{\rho_c} \Big( 2-3w+ \frac{a\Gamma}{\ch} \Big(1+\frac{\rho_x}{\rho_c}\Big) \Big)
\Big]\Big\}. \label{dcc}\ea
Thus the DM perturbations experience effectively different values of $\ch$ and $G$ due to the interactions:
\ba \frac{\ch_{\textrm{eff}}}{\ch} &=& 1-\frac{a\Gamma}{\ch}\frac{\rho_x} {\rho_c},\label{heffc}\\
\frac{G_{\textrm{eff}}}{G} &=& 1+\frac{2\rho_{\rm tot}}{3\rho_c}\frac{a\Gamma}{\ch} \frac{\rho_x}{\rho_c}\left[2-3w+\frac{a\Gamma}{\ch}\left(1+
\frac{\rho_x} {\rho_c}\right)\right].\label{geffc}\ea

The \qx model by contrast has a non-standard Euler equation (\ref{thetacprimex}), and there remains a term
proportional to $\theta_x$ which can not in general be neglected:
\ba &&\delta_c''+\ch\Big(1-2\frac{a\Gamma}{\ch}\frac{\rho_x} {\rho_c}\Big)\delta_c' =4\pi Ga^2\Big\{\rho_b\delta_b + \rho_c\delta_c \Big[1+
\nonumber\\ &&{}~
\frac{2\rho_{\rm tot}}{3\rho_c}\frac{a\Gamma}{\ch} \frac{\rho_x}{\rho_c} \Big(2-3w+\frac{a\Gamma}{\ch}\Big) \Big] \Big\}
+a\Gamma\frac{\rho_x}{\rho_c}\theta_x. \label{dcx}\ea
Nevertheless, for stable models $\theta_x$ remains small enough to be negligible and we can define the
deviations from standard growth due to the interactions via
\ba \frac{\ch_{\textrm{eff}}}{\ch} &=& 1-2\frac{a\Gamma}{\ch} \frac{\rho_x}{\rho_c},\label{heffx}\\
\frac{G_{\textrm{eff}}}{G} &=& 1+\frac{2\rho_{\rm tot}} {3\rho_c} \frac{a\Gamma}{\ch} \frac{\rho_x}{\rho_c}\left(2-3w+
\frac{a\Gamma}{\ch}\right).\label{geffx}\ea

These equations show that the differences in momentum transfer lead to a greater modification to the growth via $\ch_{\textrm{eff}}$ for the \qx model and via $G_{\textrm{eff}}$ for the \qc model, as can be seen in Fig.~\ref{effs}.
It is clear that DM perturbations in the models with large couplings are already beginning to grow exponentially at the present day (compare \cite{lop09,lop10a,lop10b,bal11b}).
In models with ${Q}_x^\mu=\Gamma {\rho}_c u_c^\mu$, as studied in \cite{val08,val10,maj10}, there is no interaction source term in the synchronous gauge version of (\ref{deltacprime}) and so the DM perturbations are stable.

\section{Conclusions}\label{conclusions}

We have studied a model of dark sector interactions with an energy transfer proportional to the DE energy density, and with momentum transfer vanishing either in the DM or the DE rest frame. We  performed an MCMC analysis
and found the best-fit parameters using a data compilation that predominantly constrains the background evolution. We found model constraints to which $\Lambda$CDM is a good fit, although
parameter degeneracies do allow for significant interaction rates at the present day and even admit the two extreme cases of zero DM at early times and zero DM today.

We analyzed the growth of structure in this model and found that the effects of large growth rates on the ISW signature in the CMB can be
suppressed by the non-standard background evolution. We also showed that interactions can greatly enhance growth in these models via effective
Hubble and Newton constants, in varying degrees depending on the momentum transfer.

There appears to be some tension between the background evolution and structure formation. The CMB, SNe and BAO data slightly favour interactions,
while the growth rate of DM perturbations likely rules out large interaction rates. There is a degeneracy with galaxy bias, which deserves further investigation.
This would
allow the use of full range of large-scale structure data and would
significantly improve the constraints on the IDE models considered
here.

Interacting models are known to be degenerate with modified gravity models~\cite{dur08,Wei:2008vw,koy09,lop10b,sim11,zia11}. It is important to break
this
degeneracy, in order to strengthen cosmological tests of GR -- currently devised tests do not incorporate the possibility of a dark sector
interaction. The key distinguishing features of IDE and modified gravity (MG) occur in: (1)~the late-time anisotropic stress, i.e. $\phi-\psi$; (2)~
the evolution of
the background DM density, $\bar\rho_c (1+z)^{-3}$; (3)~the DM-baryon bias:
 \be \nonumber
\begin{array}{lllllll}
 & & & \mbox{MG} & & & \mbox{IDE} \\
 \phi-\psi & & & \neq0 & &  & =0\\
\bar\rho_c (1+z)^{-3} & & & =\,\mbox{const} & & & \neq\,\mbox{const} \\
\delta_b-\delta_c & & & =\,\mbox{const} & & & \neq\,\mbox{const}
\\
\theta_b-\theta_c & & & =0 & &  & \mbox{can be nonzero}
\end{array}
 \ee
These features are the basis for breaking the degeneracy. For example, any difference between the metric potentials can be tested via peculiar velocities (a probe
of $\phi$), weak lensing and ISW (both sensitive to $\phi+\psi$).

\begin{table*}
\begin{center}
\begin{tabular}[c]{c|c|c|c|c|c|c|c|c|c|l}
Model &$Q_A^\mu$&$\Delta\chi^2$&$\Gamma/H_0$&$w$&$H_0$&$\Omega_bh^2$&$\Omega_ch^2$ &$n_s$&$A_s$&$\tau_{\rm rei}$\\\hline

$w=-1$ $\Gamma$CDM best-fit& \qc &$-$0.146&0.154&$-$1&70.8&0.0222&0.0974&0.959&2.16$\times10^{-9}$&0.0824\\\hline
$w=-1$ $\Gamma$CDM best-fit& \qx &$-$0.0522&0.0916&$-$1&70.0&0.0222&0.105&0.959&2.18$\times10^{-9}$&0.0852\\\hline
all $\Gamma$, all $w$ best-fit& \qc &$-$0.294&$-$0.806&$-$1.23&70.5&0.0222&0.180&0.956&2.17$\times10^{-9}$&0.0822\\\hline
all $\Gamma$, all $w$ best-fit& \qx &$-$0.0879&0.302&$-$0.951&70.1&0.0222&0.0823&0.959&2.18$\times10^{-9}$&0.0879\\\hline
all $\Gamma$, all $w$ median& \qc &-&$-$1.01&$-$1.29&70.7&0.0221&0.194&0.956&2.18$\times10^{-9}$&0.0841\\\hline
all $\Gamma$, all $w$ median& \qx &-&$-$0.578&$-$1.19&70.7&0.0222&0.164&0.958&2.18$\times10^{-9}$&0.0840\\
\end{tabular}
\end{center}
\caption{Cosmological parameters of the median and best-fit samples from CosmoMC for $w=-1$ and when the entire parameter space is
considered.\label{fits}}
\end{table*}

~\\{\bf Acknowledgments:}\\
TC is funded by a UK Science \& Technology Facilities Council (STFC) PhD studentship. KK, GZ, RM are supported by the STFC (grant no. ST/H002774/1).
RM is supported by a South African SKA Research Chair, and by a NRF (South Africa)/ Royal Society (UK) exchange grant. KK is supported by a European
Research Council Starting Grant and the Leverhulme trust. JV is supported by the Research Council of Norway. Numerical computations were done on the
Sciama High Performance Compute (HPC) cluster which is supported by the ICG, SEPNet and the University of Portsmouth.
\appendix

\section{MCMC Analysis}\label{like}
\begin{figure*}
 \includegraphics[width=\columnwidth]{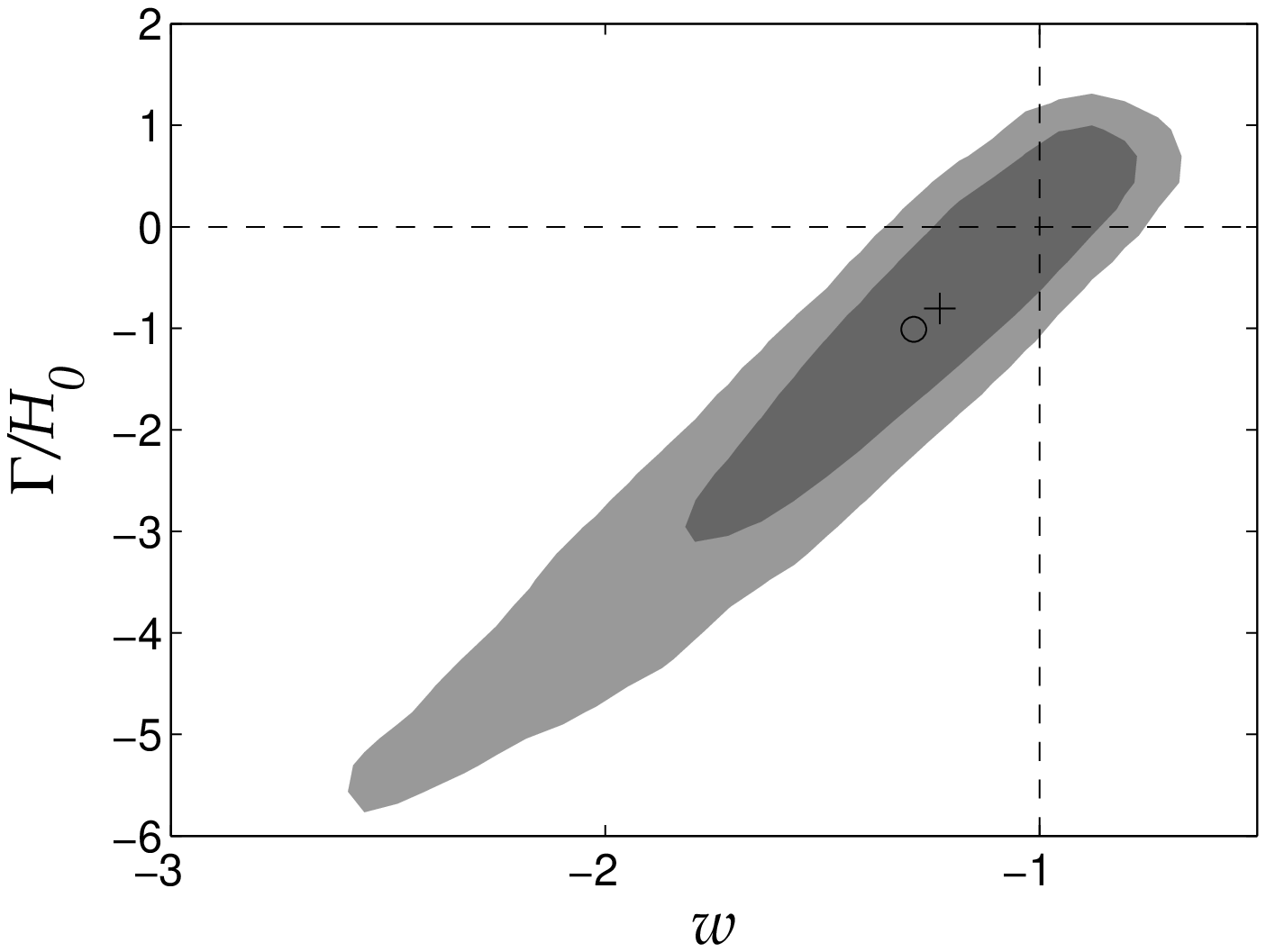}
 \includegraphics[width=\columnwidth]{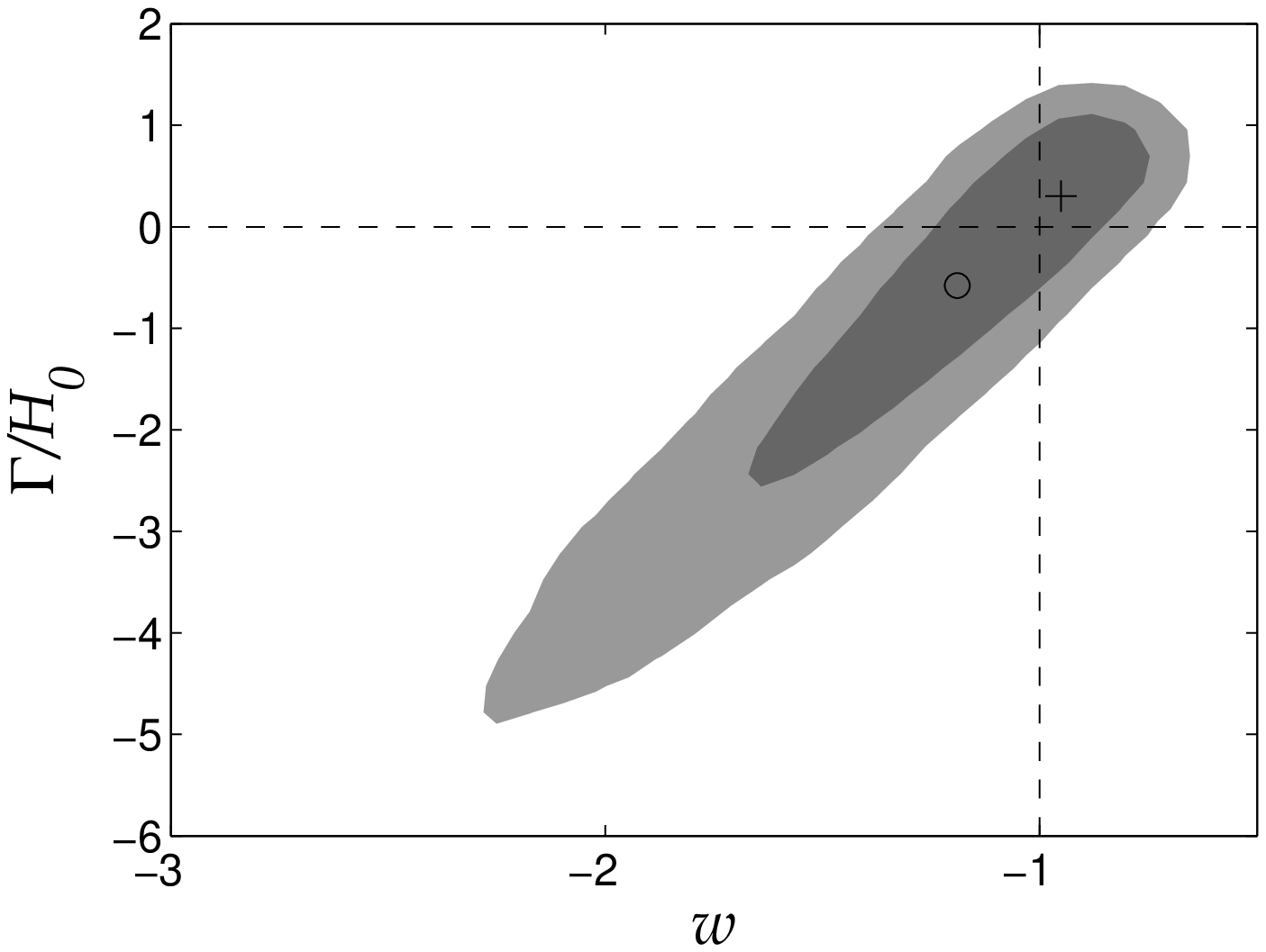}
 \caption{68\% and 95\% contours of the marginalised probability distribution for the \qc model (left) and  \qx model (right). The dashed lines cross
at the position of $\Lambda$CDM, the crosses indicate the best-fits in each case and the circles indicate the median samples. Note that some areas
appear only due to smoothing of the distributions (see Fig.~\ref{3d}).\label{complete}}
 \end{figure*}
\begin{figure*}
 \includegraphics[width=\columnwidth]{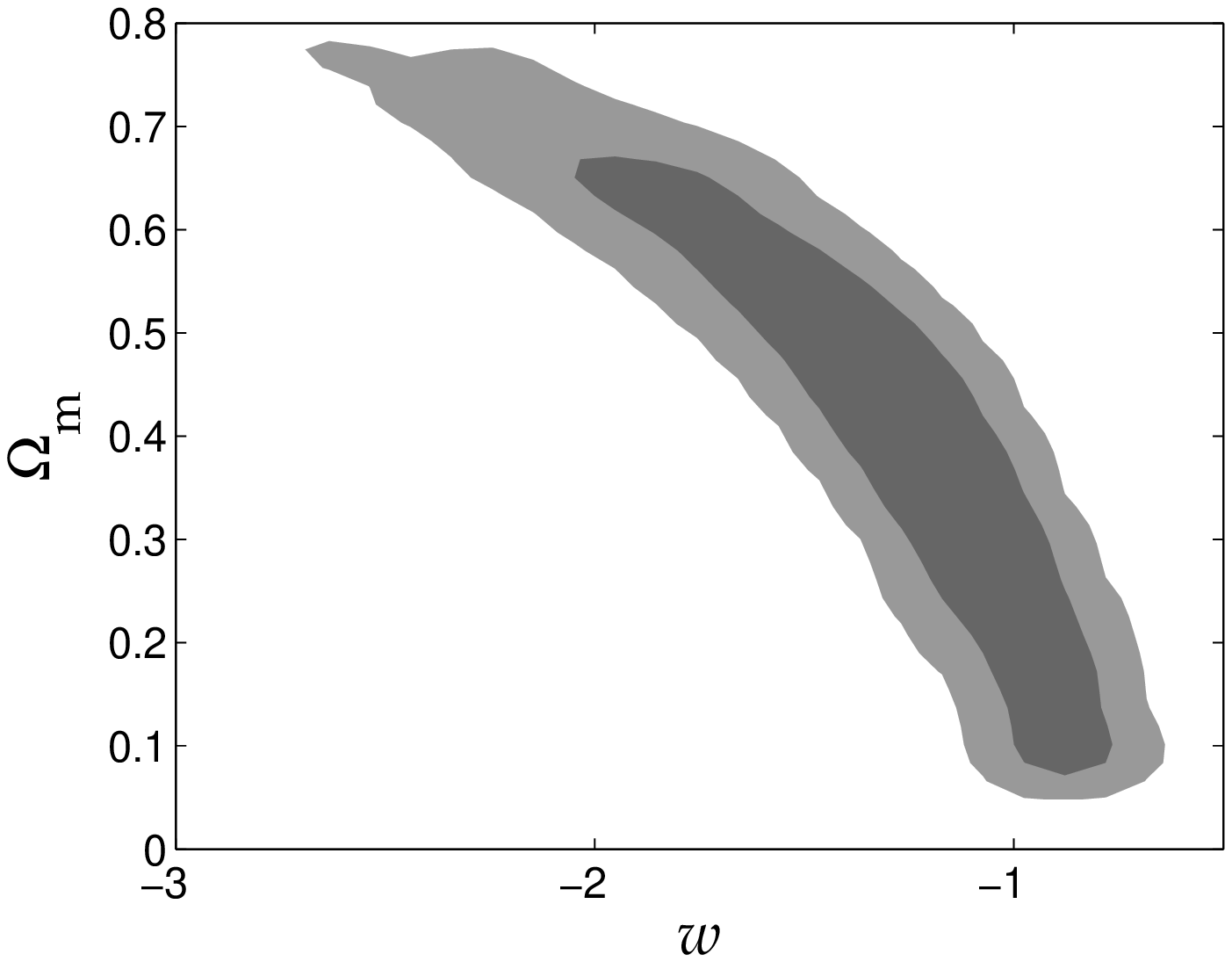}
 \includegraphics[width=\columnwidth]{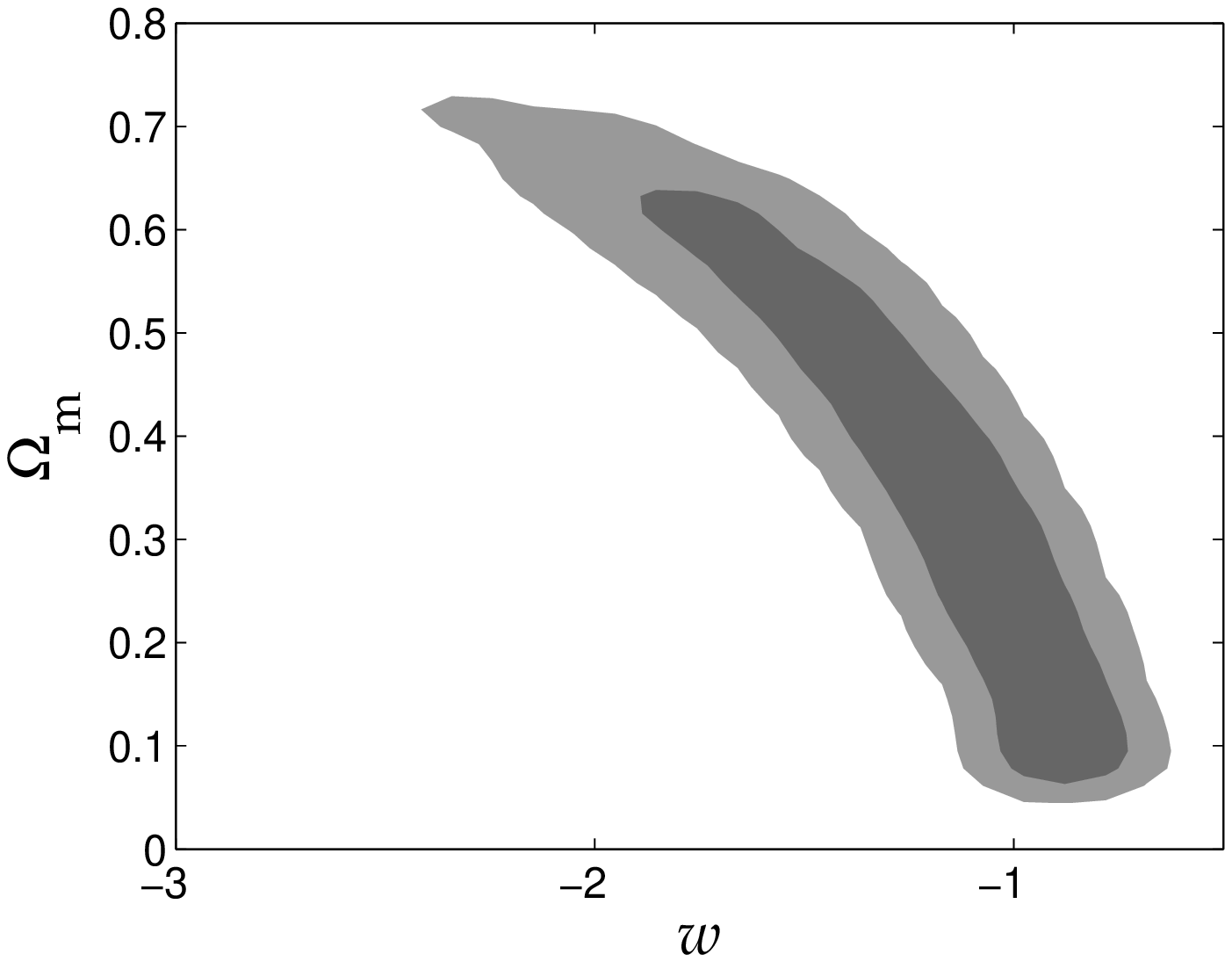}
 \caption{68\% and 95\% contours of the marginalised probability distribution in the $\Omega_m-w$ plane for the \qc model (left) and  \qx model
(right).\label{wxomegam}}

 \includegraphics[width=\columnwidth]{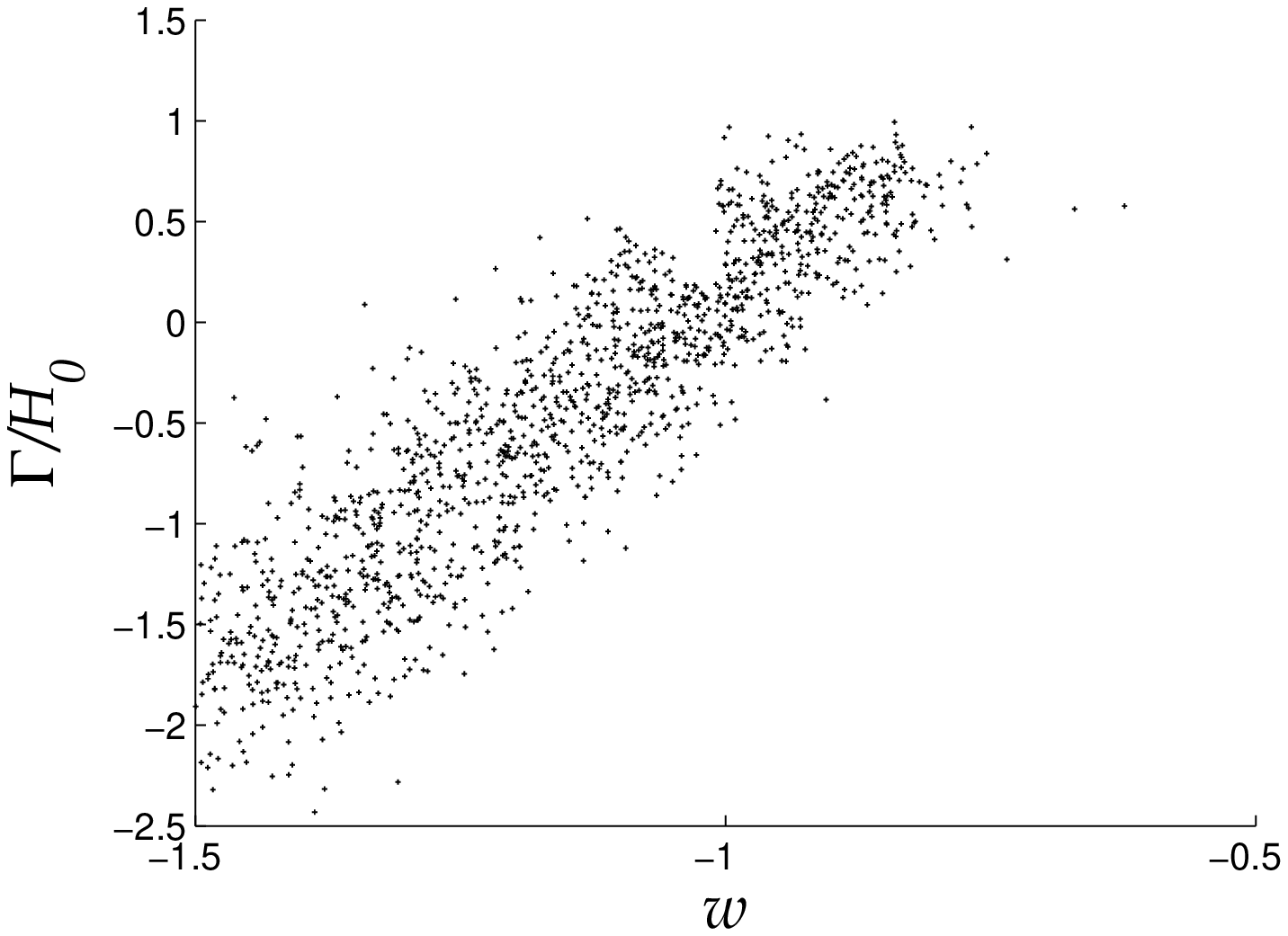}
 \includegraphics[width=\columnwidth]{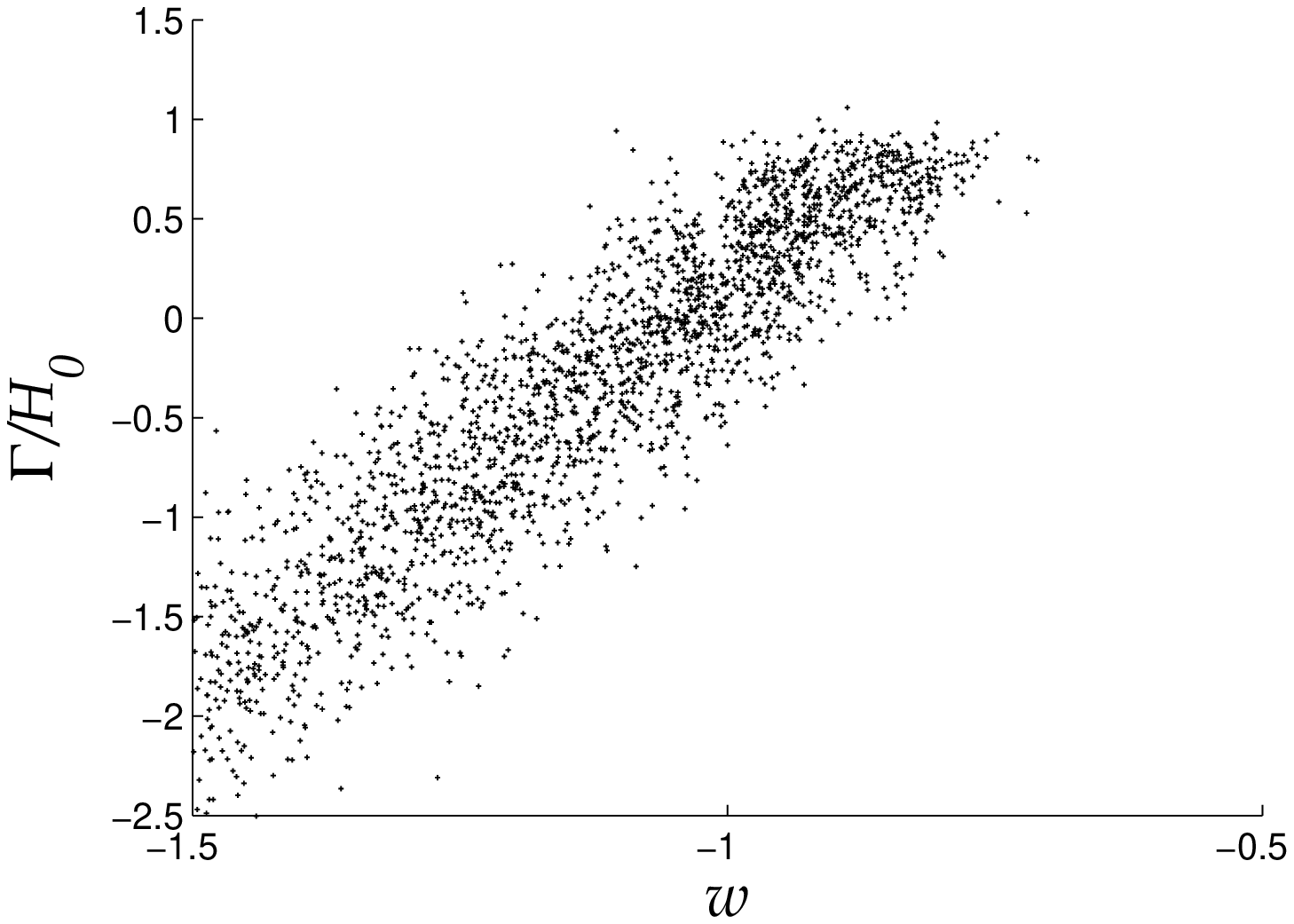}
 \caption{Distributions of accepted steps in the MCMC chains for the models with \qc (left) and \qx (right)\label{3d}}
 \end{figure*}
Using CosmoMC we explore the full parameter space of the $\Gamma w$CDM model, as illustrated in Fig.~\ref{complete}. The
code was modified to vary the two new parameters $\Gamma$ and $w$ and further coding was necessary to ensure that models with negative DM energy
densities were rejected from the MCMC analysis. In addition, the scaling solution for the background used to find the redshift of the two
BAO data points was replaced by coding to take into account the non-standard background evolution of the models. CosmoMC varies a parameter $\theta=$
100 times the ratio of the sound horizon to the angular diameter distance, in place of $H_0$, because it is more efficient. However the derivation of
$\theta$ also assumes a standard background evolution. We therefore chose not to use $\theta$, but to constrain $H_0$ directly instead. Note that the HST prior on $H_0$ assumes a particular model of $\Lambda$CDM for evolving $H(z=0.04)$ up to the present day and so has slight model dependence, which we neglect.

There is a plane of
degeneracy in the $\Gamma-w-\Omega_c$ parameter space which allows for an entire range of possibilities from zero DM at early times to zero DM at
the present day -- see Figs. \ref{complete} and \ref{wxomegam}.

The cosmological parameters of the median and best-fit models from CosmoMC for $w=-1$ and when the entire parameter space is
considered are shown in Table~\ref{fits}. For \qc models, the ISW creates a preference in the mean likelihood function
for $\Gamma<0$, as was found previously for the $Q_c^\mu= \Gamma \rho_cu_c^\mu$ models~\cite{val10}.

For \qx models, there is a preference for
$\Gamma>0$. Despite this the median samples have $\Gamma<0$ and $w<-1$ for both the \qc and \qx models. The background data therefore shows a slight
preference for values of $\Gamma<0$ and $w<-1$. Both the best-fits and the median samples however are relatively close to $\Lambda$CDM, given the
wide range of interaction strengths allowed. The $w=-1$ results
are included here to show the proximity of $\Gamma$ to 0 for these models, in line with $\Lambda$CDM.

The singularity in the perturbations at $w=-1$ leads us to impose $|1+w|<0.01$, allowing us to explore the entire
parameter space.
The effect of the $w\neq-1$ instability (\ref{instab}) is illustrated in Fig.~\ref{3d}. The wedged gaps in the
distribution of accepted MCMC chain steps are given by the boundaries of the instability region, defined by (\ref{instab2}).

\section{Initial Conditions}\label{ini}

In synchronous gauge, $\phi=B=0$ and ordinarily the residual gauge freedom is eliminated by setting $\theta_c=0$. For the \qx model the interaction term in the DM Euler equation (\ref{thetacprimex}) does not in general allow for $\theta_c=0$. However, since $\Gamma\simeq\ch_0\ll\ch$  in
(\ref{deltacprime})--(\ref{thetaxprimex}), the interactions can be neglected at early times. Using $3\psi'+k^2E'=-h/2$, where $h$ is the synchronous gauge variable \cite{ma95},
the evolution equations used to find the initial conditions for the dark sector are,
\ba&& 2\delta_c'+h'=0,~~
\theta_c'=0,\\
&&\delta_x'+3\ch(1-w)\delta_x+(1+w)\theta_x\nonumber\\
&&~{}+9\ch^2(1-w^2)\frac{\theta_x}{k^2}+(1+w) \frac{h'}{2}=0,\label{deltaxprime2}\\
&&\theta_x'-2\ch\theta_x-\frac{k^2\delta_x} {(1+w)}=0.\label{thetaxprime2}\ea
The dominant growing mode solution for $h$ found in~\cite{ma95} leads to the standard adiabatic initial conditions for DM,
\ba \delta_{c\,{\rm i}} =-\frac12h=-\frac12C(k\tau)^2,~~ \theta_{c\,{\rm i}}=0.\ea
For DE, we find the leading order solutions, in agreement with~\cite{bal10},
\ba \delta_{x\,{\rm i}} = \frac{C(1+w)k^2\tau^2}{12w-14},~~
\theta_{x\,{\rm i}} =\frac{Ck^4\tau^3}{12w-14}.\ea

\bibliography{idearticlerefs20jan}

\end{document}